 \documentclass[%
 	aps,pra,%
 	showpacs%
         ,amsmath%
         ,reprint%
         ,floatfix
         ,nofootinbib
         ,preprintnumbers%
         ]{revtex4-1}

\usepackage[%
        pdftex
	]{graphicx}

\usepackage{amssymb}

\usepackage[mathscr]{eucal}

\newcommand{\Order}{\mathcal{O}}
\newcommand{\set}[1]{\left\{ #1 \right\}}

\newcommand{\pdfrac}[2]{\frac{\partial #1}{\partial #2}}
\newcommand{\Arg}{{\rm Arg}\;}

\def\toexp{\mathop{\rm exp}}
\newcommand{\AntiTexp}{\toexp_{\rightarrow}}
\newcommand{\bra}[1]{\langle{}#1{}|}
\newcommand{\ket}[1]{|{}#1{}\rangle}
\newcommand{\bracket}[2]{\langle{}#1{}|{}#2{}\rangle}

\newcommand{\Pantiorder}{\mathop{\mathcal{P}}_{\rightarrow}}

\newcommand{\AF}{A}
\newcommand{\AD}[1]{A^{\mathrm{D} #1}}

\newcommand{\gEP}[1]{g^{(#1)}}

\newcommand{\CEP}[1]{C^{(#1)}}
\newcommand{\CReal}{\mathcal{C}}

\newcommand{\nb}{n_{\mathrm b}}

\newcommand{\ampg}{a^{(\mathrm{g})}}
\newcommand{\ampe}{a^{(\mathrm{e})}}
\newcommand{\ampLg}{a^{\mathrm{L}(\mathrm{g})}}
\newcommand{\ampLe}{a^{\mathrm{L}(\mathrm{e})}}

\newcommand{\cat}{\circ}

\begin{document}

\preprint{OCU-PHYS 385}

\title[Exotic quantum holonomy and non-Hermitian degeneracies]{%
  Exotic quantum holonomy and non-Hermitian degeneracies
  \\
  in two-body Lieb-Liniger model}

\author{Atushi Tanaka}
\homepage[]{\tt http://researchmap.jp/tanaka-atushi/}
\affiliation{ 
  Department of Physics, Tokyo Metropolitan University,
  Hachioji, Tokyo 192-0397, Japan}

\author{Nobuhiro Yonezawa}
\homepage[]{\tt http://researchmap.jp/nobuhiroyonezawa/}
\affiliation{ 
  Osaka City University Advanced Mathematical Institute (OCAMI), 
  Sumiyoshi-ku, Osaka 558-8585, Japan}  

\author{Taksu Cheon}
\homepage[]{\tt http://researchmap.jp/T_Zen/}
\affiliation{ 
  Laboratory of Physics, Kochi University of Technology, Tosa Yamada, 
  Kochi 782-8502, Japan}

\begin{abstract}
  An interplay of an exotic quantum holonomy and exceptional points is
  examined in one-dimensional Bose systems. The eigenenergy
  anholonomy, in which Hermitian adiabatic cycle induces nontrivial
  change in eigenenergies, can be interpreted as a manifestation of
  eigenenergy's Riemann surface structure, where the branch points are
  identified as the exceptional points which are degeneracy points in
  the complexified parameter space. It is also shown that the
  exceptional points are the divergent points of the non-Abelian gauge
  connection for the gauge theoretical formulation of the eigenspace
  anholonomy. This helps us to evaluate anti-path-ordered exponentials
  of the gauge connection to obtain gauge covariant quantities.
\end{abstract}


\maketitle

\section{Introduction}
\label{sec:Introduction}

A variation of a classical external parameter of a quantum system
offers a way to manipulate quantum states. Almost since the dawn of
the quantum theory, it has been recognized that the slow variation of
the parameter ensures the adiabatic time evolution, where the quantum
state can be pinned to an eigenspace of system's
Hamiltonian~\cite{Born-ZP-51-165,Kato-JPSJ-5-435}. Later, the concept
of quantum holonomy has been developed for adiabatic cycles on quantum
systems. Among them, the phase holonomy, where an adiabatic cycle
induces a nontrivial change in the phase of a quantum state~%
\cite{LonguetHiggins-PRSL-344-147,Mead-JCP-70-2284,Berry-PRSLA-392-45,%
Wilczek-PRL-52-2111}, is a textbook result
nowadays~\cite{Bohm-GPQS-2003}. Recently, the quantum holonomy of
exotic kind has been recognized in that an adiabatic cycle can induce
changes both in the eigenvalue and the eigenspace of a stationary
state~\cite{Cheon-PLA-248-285}. Such changes are also referred to as
eigenenergy and eigenspace anholonomies~\cite{Tanaka-PRL-98-160407}.

Spectral degeneracies are crucial for the quantum holonomy. The phase
holonomy is associated with a spectral degeneracy point, where a
structure mathematically identical to the magnetic monopole resides,
in the parameter space~\cite{Berry-PRSLA-392-45}. As for the exotic
quantum holonomy, it is shown, for a quantum kicked
spin-$\frac{1}{2}$, that degenerate points in the complexified
parameter space play the central role~\cite{Kim-PLA-374-1958}. Such a
non-Hermitian degeneracy point is known as an exceptional point~%
\cite{KatoExceptionalPoint,Heiss-JMP-32-3003,Heiss-CzecJP-54-1091}.
There have been considerable number of recent works on exceptional
points in non-Hermitian quantum
physics~\cite{phhqp,Heiss-CzecJP-54-1091,biorthogonal}.

It is natural to expect that the exceptional points govern the exotic
quantum holonomy in general, once we accept the view that the
anholonomies in eigenenergy and eigenspace are a manifestation of
multiple-valuedness of the solution of the eigenvalue problem in the
parameter space.  We remind the readers that an eigenvalue equation of
a Hamiltonian can be cast into an algebraic equation. Hence its
multiple-valued solutions form a family, and the family coalesces at a
branch point in the parameter space~\cite{Heiss-JMP-32-3003}.  The
encirclement of the branch point induces the permutation of the
eigenvalues and eigenspaces in the family,
%
which is to be identified as
the complex-analytic origin of the exotic quantum holonomy.  This,
however, is rather unexpected scenario, because the exceptional points
emerge only when the Hamiltonian is far from Hermitian, in spite of
the fact that the exotic quantum holonomy is induced by an adiabatic
Hermitian cycle.  Hence the question is whether 
a family
of Hermitian
Hamiltonians that define an adiabatic cycle really 
``encloses''
exceptional points.

The aim of this manuscript is to offer another example of successful
``exceptional point picture'' for the exotic quantum holonomy, in
quantum many-body systems. We examine the Lieb-Liniger model, which
describes Bose particles confined in a one-dimensional space subject
to the periodic boundary condition~\cite{Lieb-PR-130-15}. Ushveridze
showed that a non-Hermitian extension of this model has an infinite
number of exceptional points~\cite{Ushveridze-JPA-21-955}. Recently,
it is shown that the Lieb-Liniger model exhibits the eigenenergy and
eigenspace anholonomies~\cite{Yonezawa-up-20130}. Hence, our purpose
is to explain how the non-Hermitian degeneracies of this model and the
exotic quantum holonomy that occurs in Hermitian Hamiltonian is
interrelated.  We here focus on the simplest case where the number of
particles is two.  We believe that the present two-body study offers
the foundation for the case of an arbitrary number of particles.

The outline of this manuscript is the following. We introduce the
Lieb-Liniger model in Section~\ref{sec:Hermite}. We also explain that
an adiabatic Hermitian cycle of this model induces the exotic quantum
anholonomy~\cite{Yonezawa-up-20130}. We cover a non-Hermitian
extension of the Lieb-Liniger model~\cite{Duerr-PRA-79-023614} in
Section~\ref{sec:Ndim}. We outline the analytic continuation of the
quasi-momentum in Section~\ref{sec:wavenumber}. We show an
association of the eigenenergy anholonomy with exceptional points in
Section~\ref{sec:energy}. We explain the role of the exceptional
points in the gauge theoretical formulation of eigenspace anholonomy
in Section~\ref{sec:eigenspace}. We discuss the present result in
Section~\ref{sec:discussion}. We summarize this manuscript in
Section~\ref{sec:summary}.

\section{Quantum holonomy in Hermitian Lieb-Liniger 
 model}
\label{sec:Hermite}

We review the two-body Lieb-Liniger model~\cite{Lieb-PR-130-15} and
its exotic quantum holonomy~\cite{Yonezawa-up-20130}. Throughout this
manuscript, we examine the system that consists of two identical Bose
particles confined within a one-dimensional space, which is
$2\pi$-periodic. We assume that the two particles have a contact
interaction whose strength is $g$. The system is described by the
Hamiltonian:
\begin{equation}
  \label{eq:defH}
  H(g) 
  = -\frac{1}{2}\left(\pdfrac{{}^2}{x_1^2}+  \pdfrac{{}^2}{x_2^2}\right)
  + {g}\delta(x_1 - x_2)
  ,
\end{equation}
where the units are chosen such that $\hbar$ and the mass of a Bose
particle are $1$.

We explain the standard method to solve the eigenvalue problem of
$H(g)$~\cite{Lieb-PR-130-15}.  We employ the Bethe ansatz, where an
eigenfunction is expressed by two plane waves that are associated with
quasi-momentum (rapidity) $k_j$ ($j=1,2$).  The total momentum
$\bar{k}\equiv k_1 + k_2$ must be an integer, because the periodic
boundary condition is imposed. On the other hand, the difference in
the quasi-momenta $k\equiv k_2-k_1$ satisfies the condition that
\begin{equation}
  \label{eq:defJ}
  J(g, k)
  \equiv k + \frac{2}{\pi}\arctan \frac{k}{g} 
  ,
\end{equation}
is an integer~\cite{Lieb-PR-130-15}. Throughout this paper, $\arctan$
denotes the principal value of the inverse tangent. For odd $J(g,k)$, 
$k$ satisfies
\begin{equation}
  \label{eq:Bethe21g}
  k/g
  = \cot(\pi k/2)
  ,
\end{equation}
whereas even $J(g, k)$ implies
\begin{equation}
  \label{eq:Bethe21e}
  k/g = -\tan(\pi k/2)
  .
\end{equation}
We call \eqref{eq:Bethe21g} and~\eqref{eq:Bethe21e} the Bethe
equations.

We look for the solution of the Bethe equations for real $g$.  Let
$k_{n}(g)$ denote the solution that satisfies $k_{n}(0)=n$, and
smoothly depends on $g$ in the real axis $-\infty < g < \infty$.  It
suffices to examine the case where $n$ is a non-negative integer.
$k_{n}(g)$ is either real or pure imaginary.  The latter case
describes the ``clustering'' of two particles, which occurs only when
$n=0$ and $g<0$, or, $n=1$ and $g < -2/\pi$.  We here summarize the
relevant facts on $k_{n}(g)$ shown in Ref.~\cite{Lieb-PR-130-15}: (a)
$k_{n}(\infty)=n+1$, and accordingly $J(k_n(g),g)=n+1$ for $g>0$; (b)
For $n>1$, $k_{n}(-\infty)=n-1$ and accordingly $J(k_n(g),g)=n-1$ for
$g<0$; (c) $k_0(g)$ and $k_1(g)$ satisfy \eqref{eq:Bethe21g} and
\eqref{eq:Bethe21e}, respectively.

There remain freedoms to choose the signs of $\Im k_n(-\infty)$ for
$n=0$ and $1$.  Here we carry out the analytic continuation of
$k_n(g)$ through the lower half plane of $g$ from $g>0$. Although this
choice is arbitrary for our purpose, the present choice is consistent
with the condition that the non-Hermitian Lieb-Liniger model describes
the (forward-)time evolution correctly (see, Sec.~\ref{sec:Ndim}).  As
a result, we obtain $k_n(-\infty) = -i\infty$ for $n=0$ and $1$.

The eigenstates of the two-body Lieb-Liniger model are specified by
two quantum numbers $\bar{k}$ and $n$, which must be both even or odd.
The corresponding eigenenergy is
\begin{equation}
  E_{\bar{k},n}(g) 
  = \frac{1}{2}\left({\bar{k}}^2 + \left\{k_n(g)\right\}^2\right)
  .
\end{equation}

We introduce a cycle $\CReal(g_0)$ in real $g$-space to investigate
the exotic quantum holonomy.  The initial point of $\CReal(g_0)$ is
$g_0$. We increase $g$ adiabatically during $g_0 \le g <
\infty$. Then, $g$ is suddenly flipped from $\infty$ to $-\infty$.
Such a sudden flip has been investigated both in
theory~\cite{Olshanii-PRL-81-938} and
experiments~\cite{Haller-Science-325-1224,Haller-PRL-104-153203} to
approach super Tonks-Girardeau gas.  To finish $\CReal(g_0)$, $g$ is
adiabatically increased from $-\infty$ to $g_0$.

The exotic quantum holonomy is found in 
the
two-body Lieb-Liniger model
along the cycle $\CReal(g_0)$~\cite{Yonezawa-up-20130}. We initialize
the interaction strength as $g=g_0$, and prepare the system to be in
the eigenstate specified by quantum numbers $(\bar{k},n)$.  As we
increase $g$ adiabatically, the energy of the system follows
$E_{\bar{k},n}(g)$, which increases monotonically.  When we arrive
$g=\infty$, we assume that we suddenly switch the value of $g$ from
$\infty$ to $-\infty$, keeping the system remain unchanged.  Because
of $E_{\bar{k},n}(\infty)=E_{\bar{k},n+2}(-\infty)$, the system is in
the $(\bar{k},n+2)$ state after the switch.  As we increase $g$ from
$-\infty$ to $g_0$ adiabatically, the energy of the whole system
arrives at $E_{\bar{k},n+2}(g_0)$, which does not agree with the
initial energy. This is the eigenenergy anholonomy of the two-body
Lieb-Liniger model.  Because the Hamiltonian is Hermitian for real
$g$, the eigenenergy anholonomy implies the eigenspace anholonomy,
i.e., the initial and final state vectors correspond to different
eigenenergies and are thus orthogonal.

\section{Non-Hermitian Lieb-Liniger model}
\label{sec:Ndim}

We shall show, in the following sections, that the spectral
degeneracies that are hidden in the complexified parameter space
governs the exotic quantum holonomy of the two-body Lieb-Liniger
model.  To carry out this, we introduce the complexification of the
coupling strength $g$. We outline formal aspects of the consequence
of the complexification in this section, and the details of the
analytic continuations of relevant quantities will be explained in the
following sections. We refer Ref.~\cite{biorthogonal} for the theory
of non-Hermitian eigenvalue problem.

Here we focus on the lower-half plane of $g$.  This is just a matter
of convention in our analysis since there is a symmetry about the real axis
in the complex $g$-plane.  However, the present choice is
suitable once we realize that the complexified Lieb-Liniger model
describes the dissipative effect. In Ref.~\cite{Duerr-PRA-79-023614},
it is shown that the presence of inelastic collisions implies that the
imaginary part of the effective one-dimensional coupling constant must
be zero or negative, i.e., $\Im (g) \le 0$.

First of all, the complexification makes the Lieb-Liniger Hamiltonian
$H(g)$~~\eqref{eq:defH} non-Hermitian.  An immediate consequence is
that the two eigenvalue problems for $H(g)$ and
$\left[{H}(g)\right]^{\dagger}$ become different because of the
relation $\left[{H}(g)\right]^{\dagger} = H(g^*)$.  An eigenvalue of
$\left[{H}(g)\right]^{\dagger}$ is given by 
$\left[E_{\bar{k}, n}(g)\right]^*$, which may not be identical to 
$E_{\bar{k}, n}(g)$.

From the comparison of the spectrum sets of ${H}(g)$ and
$\left[{H}(g)\right]^{\dagger}$, there is a unique pair $(n',n)$ that
satisfies
\begin{equation}
 \left\{E_{\bar{k}, n'}(g^*)\right\}^* = E_{\bar{k}, n}(g) 
 ,
\end{equation}
where we assume that ${H}(g)$ has no spectral degeneracy within the
subspace specified by the total momentum $\bar{k}$. This assumption
holds in the vicinity of the real axis.  We note that the
correspondence between $n'$ and $n$ generally depends on the details
of the analytic continuation and the value of $g$.  We ignore the case
$n'\ne n$ because we focus on the neighborhood of the real axis of
$g$, and assume 
$\left\{E_{\bar{k}, n}(g^*)\right\}^* = E_{\bar{k}, n}(g)$. This implies
\begin{equation}
  \label{eq:kBiorthogonalSymmetry}
  \left\{k_n(g^*)\right\}^*=s k_n(g)
  ,
\end{equation}
where $s$ is either $1$ or $-1$, depending on how the analytic
continuation of $k_n(g)$ is carried out. Also, $s$ may depend on $n$
and $g$.

Eigenfunctions of $H(g)$ can be obtained through a standard way with
the help of the Bethe ansatz. We refer Ref.~\cite{Lieb-PR-130-15} for
details.  Let $\psi_{\bar{k}, n}(x_1, x_2)$ be an eigenfunction that
corresponds to the eigenenergy $E_{\bar{k}, n}(g)$.  Because of the
Bose statistics, we will write down only the expressions of
eigenfunctions in the region 
$R_1\equiv\set{(x_1, x_2)|0 \le x_1 \le x_2 \le L}$:
\begin{equation}
  \label{eq:defPsi}
  \psi_{\bar{k}, n}(x_1, x_2;g)
  = \Phi_{\bar{k}}\left(\frac{x_1+x_2}{2}\right)\Psi_{n}(x_2-x_1; g)
  ,
\end{equation}
where
\begin{eqnarray}
  \Phi_{\bar{k}}(X)
  &
  \equiv
  &
  \frac{1}{\sqrt{2\pi}} e^{i \bar{k} X}
  ,\\
  \Psi_{n}(x; g)
  &
  =
  &
    \begin{cases}
    \frac{1}{\sqrt{2\pi}} 
    \ampg(k)
    \cos\left[k (x-\pi)/2\right]
    & 
      \text{%
      for even $n$,
    }
    \\
    \frac{1}{\sqrt{2\pi}} 
    \ampe(k)
    \sin\left[k (x-\pi)/2\right]
    & 
      \text{%
      for odd $n$,
    } 
    \\
    \end{cases}
\end{eqnarray}
and the following abbreviation is introduced:
\begin{equation}
  \label{eq:kabb}
  k\equiv k_{n}(g) 
  .
\end{equation}
Here we choose the normalization constants $\ampg(k)$ and $\ampe(k)$
being independent of $n$ in order to ensure that the eigenfunctions
are continuous with respect to $g$.

We turn to $\psi^{\mathrm{L}}_{\bar{k},n}(x_1, x_2; g)$ that is an
eigenfunction of $\left[{H}(g)\right]^{\dagger}$ corresponding to the
eigenvalue $\left\{E_{\bar{k}, n}(g)\right\}^*$. Because of
$\left[{H}(g)\right]^{\dagger} = {H}(g^*)$, we find
\begin{equation}
  \label{eq:defPsiL}
  \psi^{\mathrm{L}}_{\bar{k}, n}(x_1, x_2;g)
  = \Phi_{\bar{k}}\left(\frac{x_1+x_2}{2}\right)
  \Psi^{\mathrm{L}}_{n}(x_2-x_1; g)
  ,
\end{equation}
where
\begin{equation}
  \Psi^{\mathrm{L}}_{n}(x; g)
  =
    \begin{cases}
    \frac{1}{\sqrt{2\pi}} 
    \ampLg(\tilde{k})
    \cos\left[\tilde{k}(x-\pi)/2\right]
    & 
      \text{%
      for even $n$,
    }
    \\
    \frac{1}{\sqrt{2\pi}} 
    \ampLe(\tilde{k})
    \sin\left[\tilde{k}(x -\pi)/2\right]
    & 
      \text{%
      for odd $n$,
    } 
    \\ 
    \end{cases}
\end{equation}
and we introduce another abbreviation 
\begin{equation}
  \tilde{k}\equiv k_{n}(g^*)
  .
\end{equation}
We choose the following normalization condition
\begin{equation}
  \bracket{\psi^{\mathrm{L}}_{\bar{k}', n'}(g)}{\psi_{\bar{k}, n}(g)}
  = \delta_{\bar{k}',\bar{k}}\delta_{n',n}
  .
\end{equation}
This implies
\begin{eqnarray}
  \left\{\ampLg(\tilde{k})\right\}^*\ampg(k)
  &
  =
  &
  2
  \left(
    1 + \frac{\sin(\pi k)}{\pi k}\right)^{-1}
  ,\\
  \left\{\ampLe(\tilde{k})\right\}^*\ampe(k)
  &
  =
  &
  2 s
  \left(
    1 - \frac{\sin(\pi k)}{\pi k}\right)^{-1}
  ,
\end{eqnarray}
where $s$ was introduced in \eqref{eq:kBiorthogonalSymmetry}.

\section{Quasi-momentum Riemann surface}
\label{sec:wavenumber}

The exotic quantum holonomy in Lieb-Liniger model can be cast into
the anholonomy of the quasi-momentum $k_n(g)$. As explained in
Sec.~\ref{sec:Hermite}, the cycle $\CReal(g_0)$ changes $k_n(g_0)$
into $k_{n+2}(g_0)$. We here examine the analytic continuation of
$k_n(g)$ to provide the basis of the analysis in the following
sections.

The quasi-momenta $k_n(g)$'s in the complex $g$-plane form Riemann
surfaces.  There are two kinds of Riemann surfaces that correspond
to $k_n(g)$'s with even and odd $n$s, respectively, of the two-body
Lieb-Liniger model.  We first look at the Riemann surface that
involves $k_n(g)$'s with even $n$.

We carry out the analytic continuation of $k_n(g)$.  When $g$ is
positive and $\epsilon$ is sufficiently small,
$J(g+\epsilon,k_n(g+\epsilon))=n+1$ holds.  Expanding this equation
with respect to 
the
small parameter $\epsilon$, we obtain
\begin{equation}
  \label{eq:kIncrement}
  k_n(g+\epsilon) 
  = k_n(g) 
  + \epsilon G(g, k_n(g))
  + \mathcal{O}(\epsilon^2)
  ,
\end{equation}
where
\begin{equation}
  G(g,k)
  \equiv
  -\frac{\partial_g J(g,k)}%
  {\partial_k J(g,k)}
  .
\end{equation}
\eqref{eq:kIncrement} is applicable to negative $g$ and arbitrary $n$.
Note that \eqref{eq:kIncrement} makes sense only when the denominator
of $G(g,k)$ is non-zero.  As long as we can find a way to avoid the
breakdown of this condition, \eqref{eq:kIncrement} provides a way to
obtain $k_n(g)$ with an arbitrary $g$.

We examine the condition that the procedure above is inapplicable.  We
assume that $(g, k_n(g))$ satisfies $\partial_k J(g,k) = 0$. This
implies that $g$ is a branch point of $k_n(g)$. Indeed, we obtain
\begin{equation}
  \label{eq:kIncrement2}
  k_n(g+\epsilon) 
  = k_n(g) 
  \pm \sqrt{\epsilon G^{(2)}(g,k_n(g))}
  + \mathcal{O}(\epsilon)
  ,
\end{equation}
where
\begin{equation}
  G^{(2)}(g,k)
  \equiv
  -\frac{2\partial_g J(g,k)}{\partial_k^2 J(g,k)}
  ,
\end{equation}
as long as the denominator of $G^{(2)}(g,k)$ is nonzero. When
$\partial_k^2 J(g,k)$ vanishes, the branch point is of third or 
higher
order. For the two-body Lieb-Liniger model, the degree of all branch
points is two, as is to be seen below.

We enumerate the branch points by solving the equation $\partial_k
J(g,k) = 0$ under the condition that $J(g,k)$ is an integer.  This is
equivalent to
\begin{equation}
  \label{eq:EP_eq}
  k = \pm \sqrt{-g(g + 2/\pi)}
\end{equation}
as long as $|g| < \infty$
holds.

In the real axis, the family $k_n(g)$'s with even $n$ has only one
branch point at $g=0$, where two quasi-momenta $k_0(g)$ and $-k_0(g)$
degenerate and the other quasi-momenta are not involved.  Accordingly
the real branch 
point does
not involve any spectral degeneracy.

\begin{figure}[ht]
  \centering

  \includegraphics[width=0.46\textwidth]{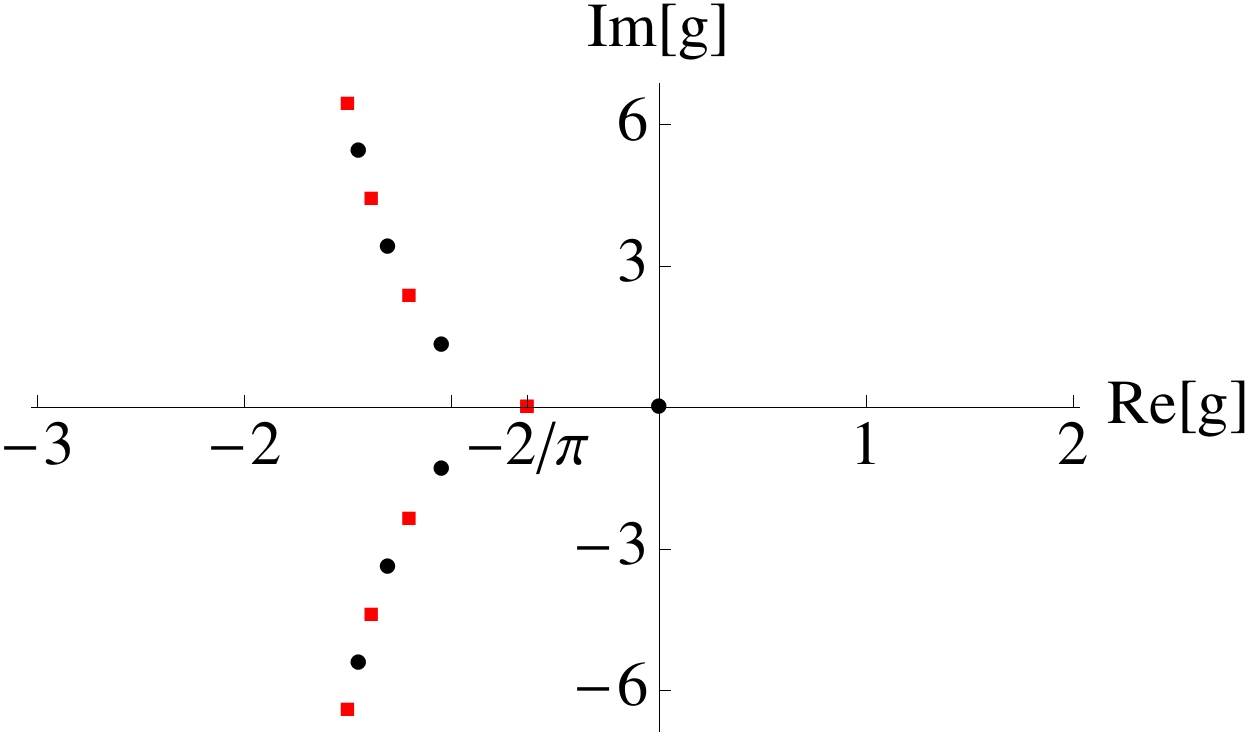} 

  \caption{%
    (Color online)
    Branch points of $k_n(g)$ in complex 
    $g$-plane.
    Circles and squares correspond to even and odd $n$, respectively.
    When $n$ is even, we numerically obtain $g$ that
    simultaneously satisfies 
    \eqref{eq:EP_eq} and~\eqref{eq:Bethe21g}.
    There is only a 
    single
    branch point
    $g=0$ on the real axis.
    All other complex branch points are in the region
    $\Re\; g < 0$.
    We obtain the branch points for the 
    odd
    $n$ case by
    solving 
    \eqref{eq:EP_eq} and~\eqref{eq:Bethe21e}.
    There is only a 
    single
    branch point
    $g=-2/\pi$ on the real axis.
    All other complex branch points are in the region
    $\Re\; g < -2/\pi$.
  }
  \label{fig:BranchPoints}
\end{figure}

Complex branch points of quasi-momenta, which are obtained
numerically, are depicted in Figure~\ref{fig:BranchPoints}.  All the
complex branch points are confined in the region $\Re\; g < 0$.  These
complex branch points involve spectral degeneracies, because two
quasi-momenta that are degenerate at a complex branch point provide
different eigenenergies, except at the branch point.  This leads to a
coalescence of the eigenspaces.  Such complex branch points are called
Kato's exceptional
points~\cite{KatoExceptionalPoint,Heiss-JMP-32-3003,Heiss-CzecJP-54-1091}.

Our numerical result is consistent with the argument above in the
sense that all complex branch points are of degree two. We find that a
complex branch point always involves $k_0(g)$, the quasi-momentum of
the ground state.  As far as we see, there is no branch point that
involves two ``excited states'' $k_n(g)$ and $k_{n'}(g)$ with $n,n'>1$
at the same time.  Also, each $k_n(g)$ with $n>1$ is involved with a
complex branch point, which is denoted by $g^{(n)}$.  Namely, $k_0(g)$
and $k_n(g)$ degenerate at $(g,k)=(g^{(n)},k^{(n)})$, where $k^{(n)}
\equiv k_0(g^{(n)}) = k_n(g^{(n)})$.

The numerical result can be explained qualitatively by a perturbation
expansion around $g=-\infty$~\cite{Ushveridze-JPA-21-955}. The
quasi-momentum of the ground state diverges as $k_0(g) \sim i g$ at
$g\sim-\infty$~\cite{Lieb-PR-130-15}, where we ignore the small
$g^{-1}$ correction.  On the other hand, the quasi-momentum of the
$n$-th excite state ($n>1$) converges to a constant value, i.e.,
$k_n(g) \sim n-1$ at $g\sim-\infty$~\cite{Lieb-PR-130-15}.  Hence
$k_0(g)$ and $k_n(g)$ coincides at $g\sim -i (n-1)$, which is
considered to be an approximation of $g^{(n)}$ in the lower
half-plane.  Although this argument is consistent only when $n$ is
large enough, the present estimation seems to be applicable to smaller
$n$'s (see figure~\ref{fig:BranchPoints}).  We also remark that the
configuration of exceptional points in the upper half-plane is due to
the other branch of $k_0(g)$, i.e., $k_0(g)\sim -i g$ at 
$g\sim-\infty$.

\begin{figure*}[ht] 
  \centering

  \includegraphics[width=0.28\textwidth]{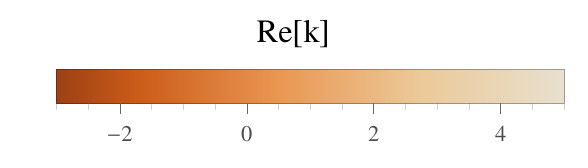}
  \hspace{1em}
  \includegraphics[width=0.28\textwidth]{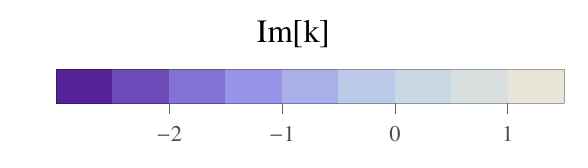}
  \\[1.5\baselineskip]

  \includegraphics[width=0.28\textwidth]{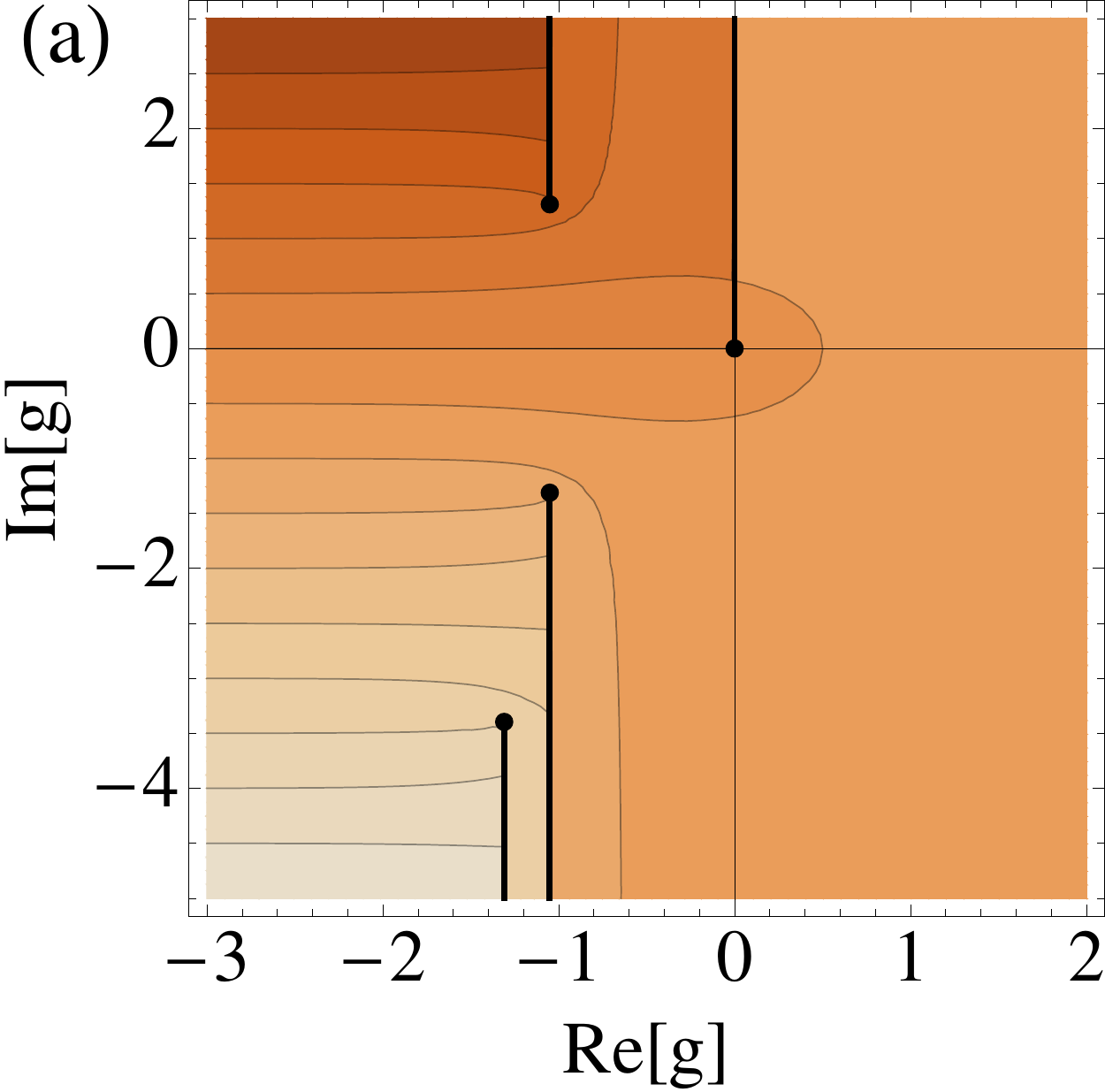}
  \hspace{1em}
  \includegraphics[width=0.28\textwidth]{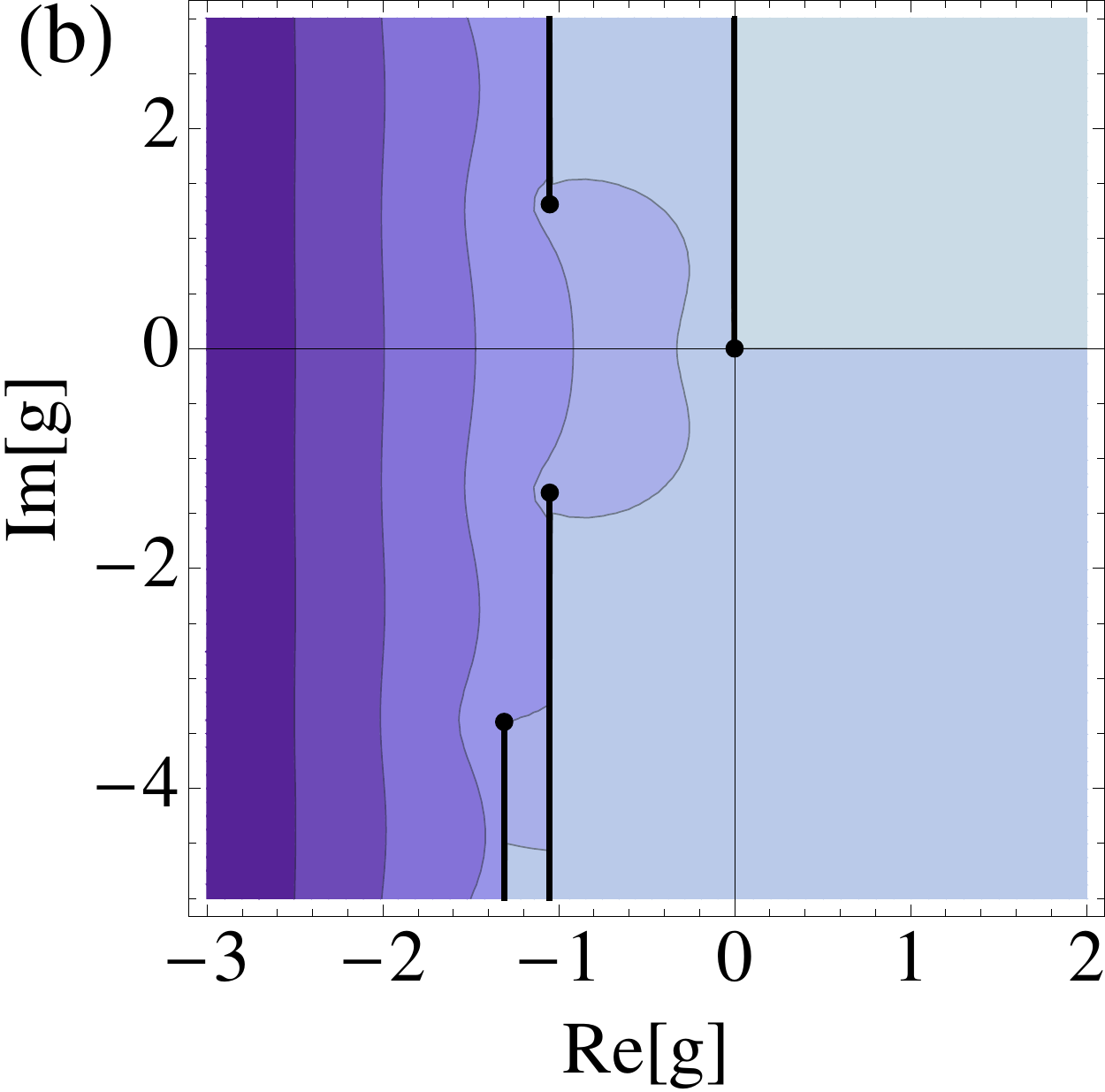}    
  \\[1.5\baselineskip]

  \includegraphics[width=0.28\textwidth]{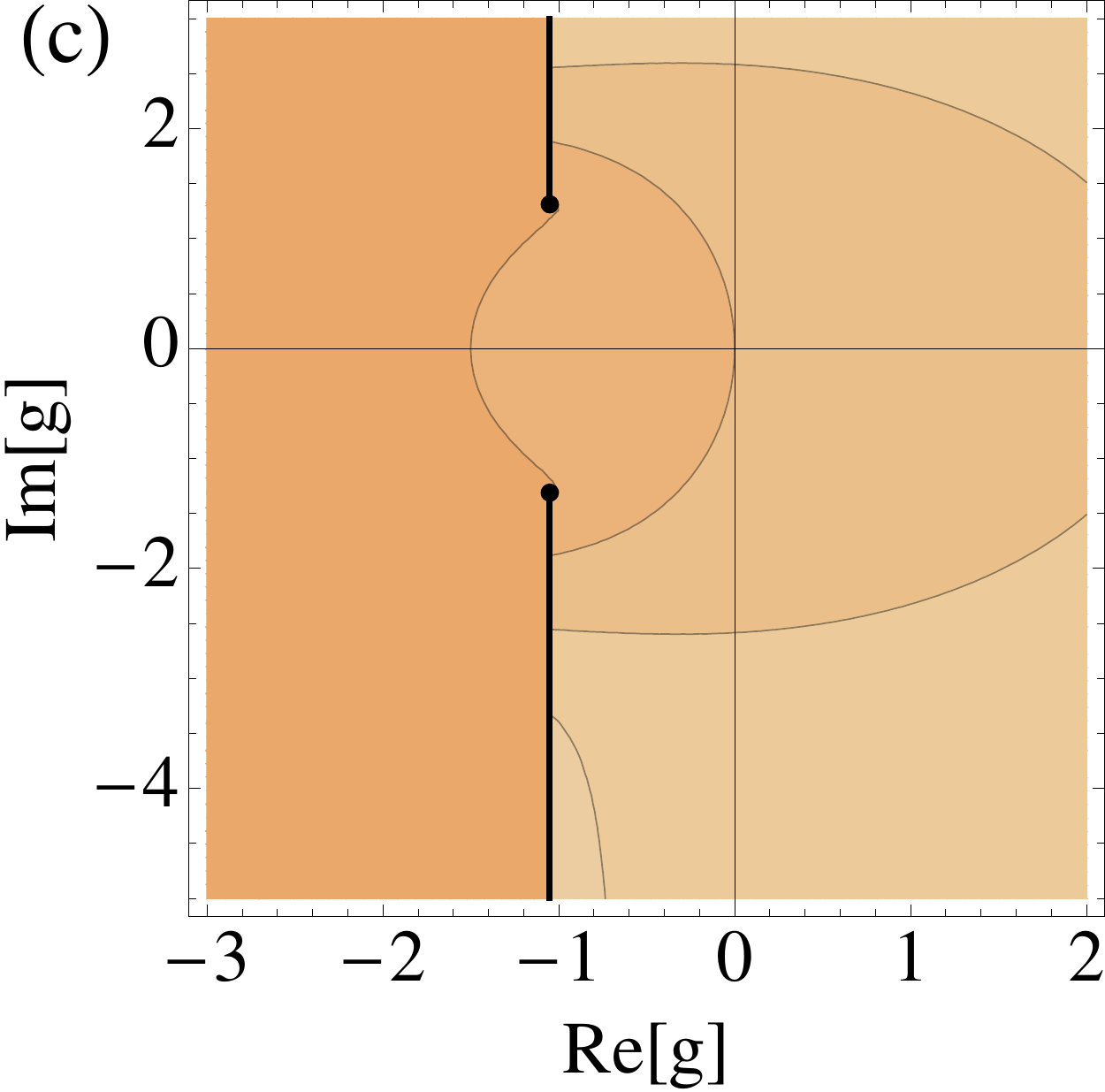}
  \hspace{1em}
  \includegraphics[width=0.28\textwidth]{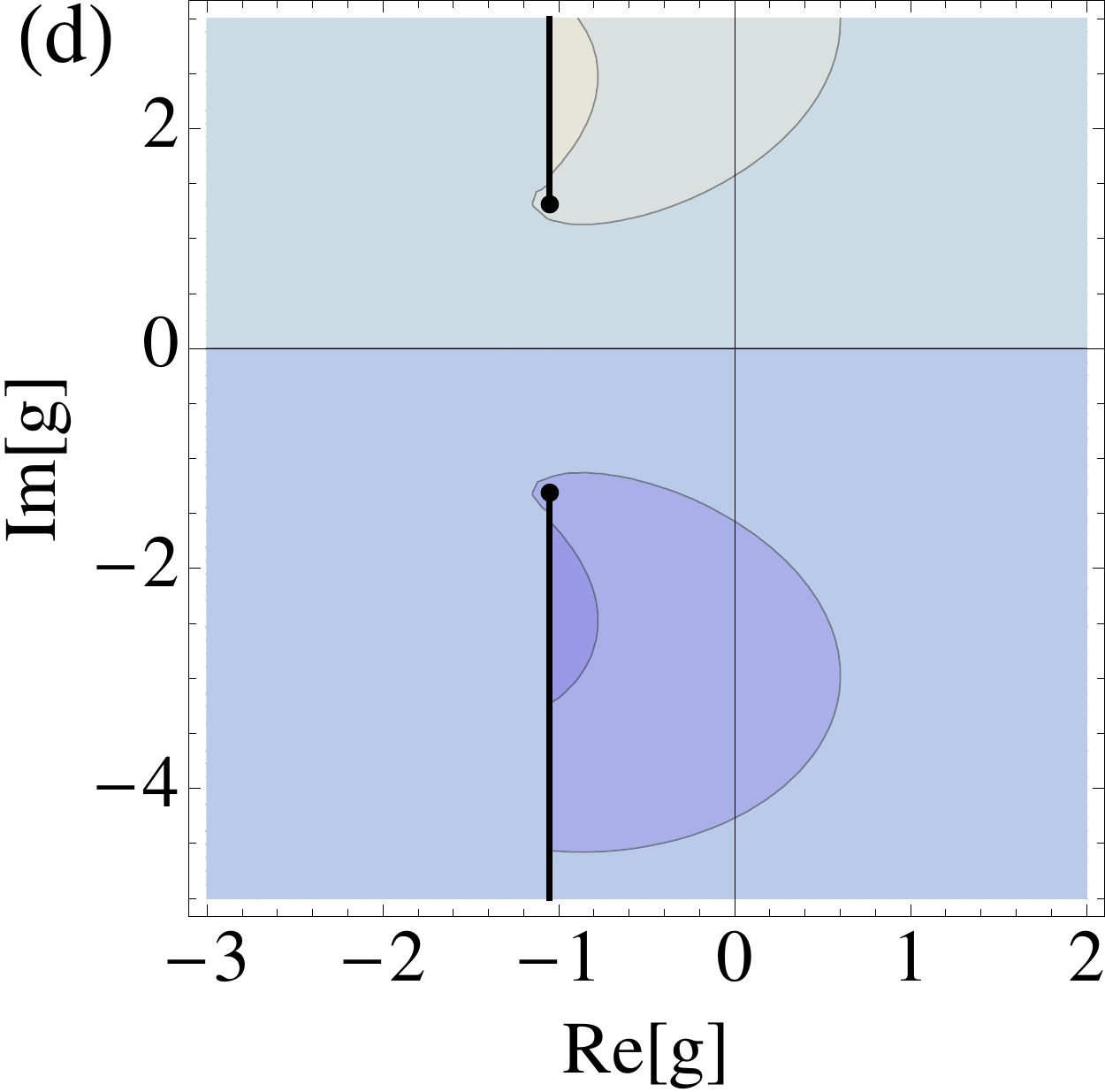}
  \\[1.5\baselineskip]

  \includegraphics[width=0.28\textwidth]{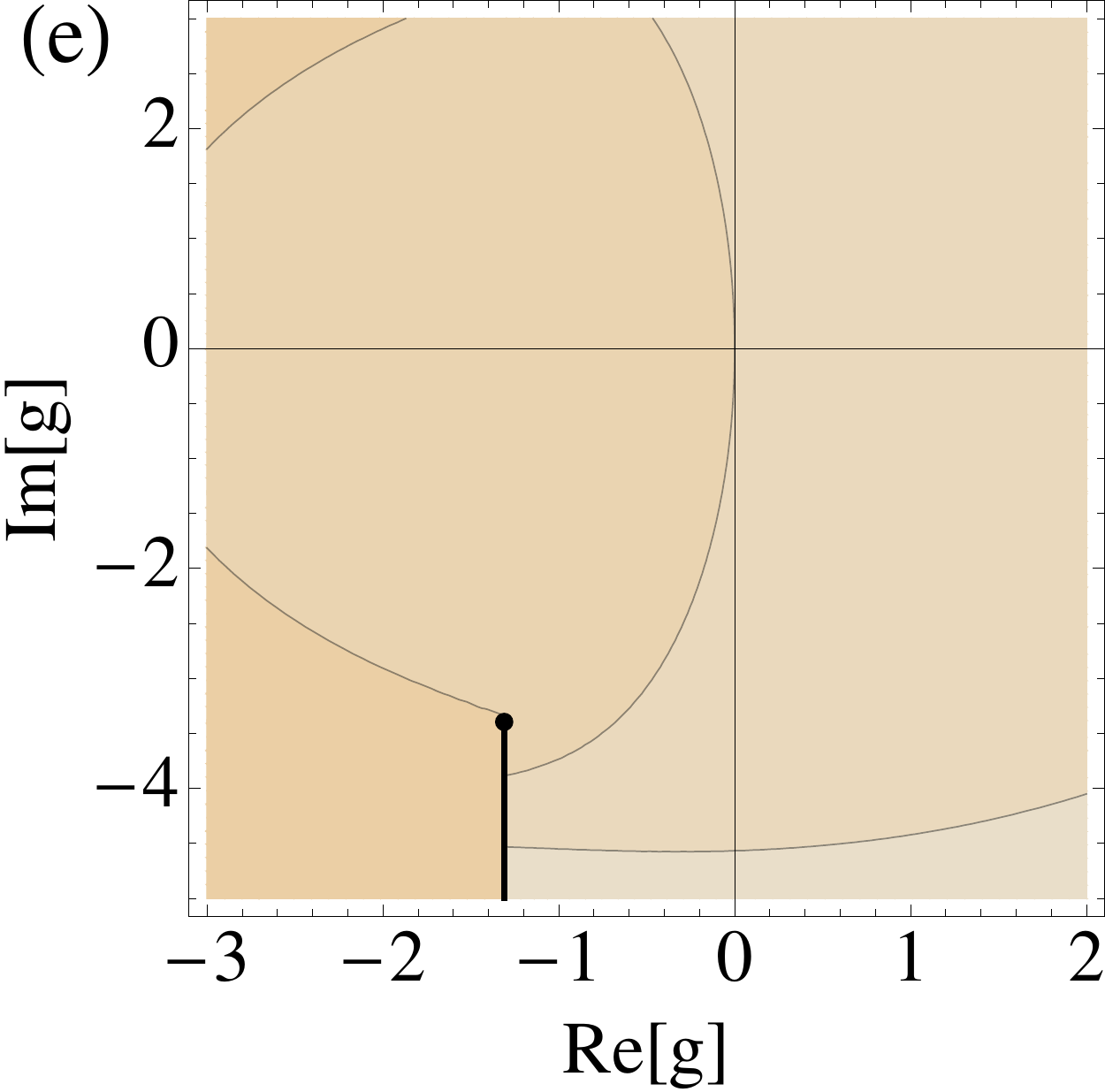}
  \hspace{1em}
  \includegraphics[width=0.28\textwidth]{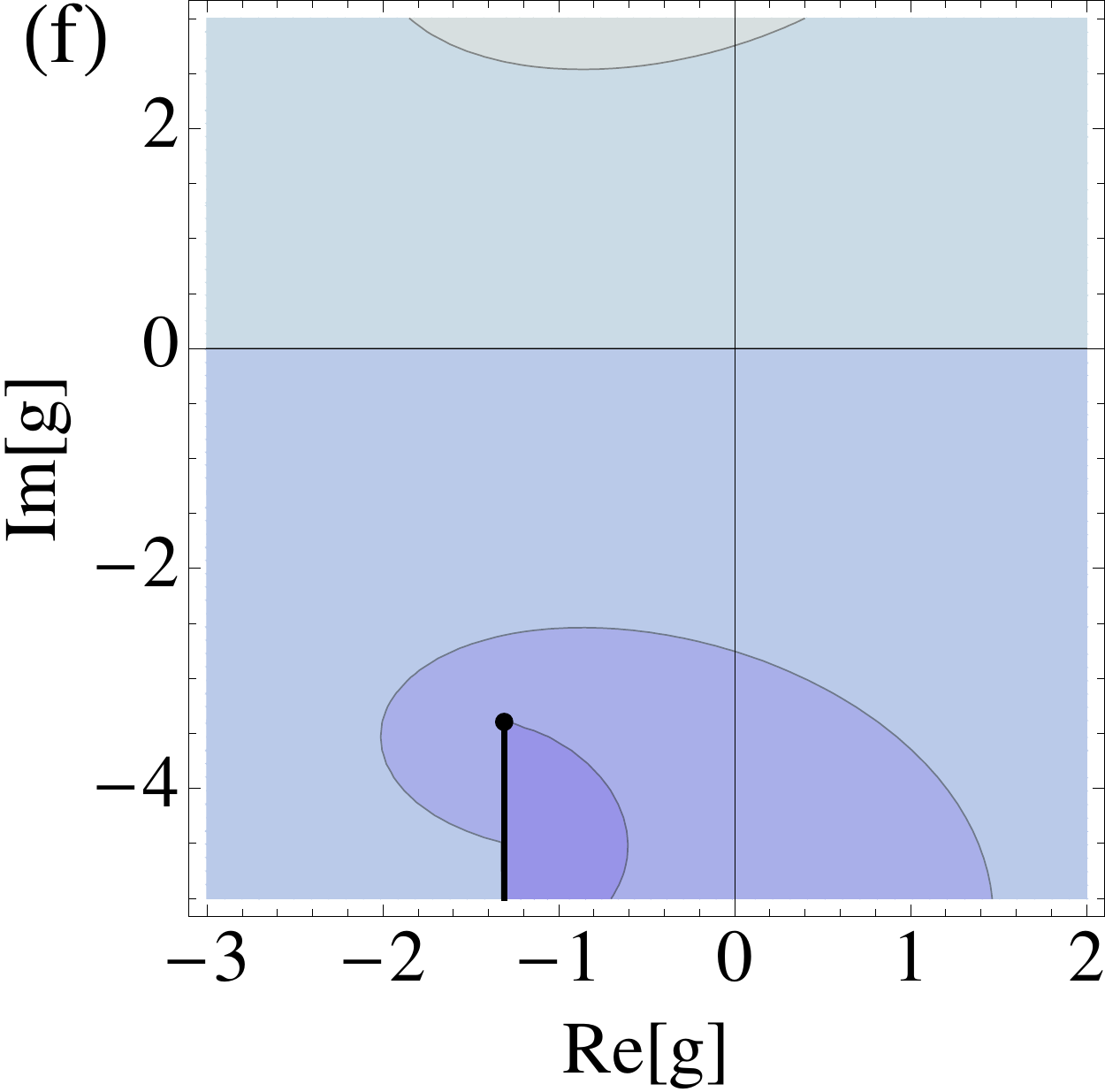}
  \\[1.5\baselineskip]

  \caption{%
    The Riemann sheets $k_n(g)$: (1st row) $n=0$; (2nd row) $n=2$;
    (3rd row) $n=4$.  The real and imaginary parts are shown in the
    left and right columns, respectively. While all exceptional
    points appear in $k_0(g)$, each $k_n(g)$ ($n>1$) has a single
    exceptional point.  Bold lines indicate branch cuts.  These
    Riemann sheets are interconnected by the branch cuts in the lower
    half plane. This is due to our choice of the sign of ``bound
    states'' $k_{0}(g)$ and $k_{1}(g)$ in the real axis.  }
  \label{fig:EPs_in_even_sheets}
\end{figure*}

We construct a Riemann sheet $k_n(g)$ in the complex plane as follows.
For real numbers $g'$ and $g''$, $k_n(g'+i g'')$ is extended from
$k_n(g')$ along the line parallel to the imaginary axis, if there is
no branch point in the interval between $g'$ and $g'+i g''$. Branch
cuts are chosen to be parallel to the imaginary axis. We also require
that the branch cuts do not traverse the real axis.  When there is a
branch point in the real axis, the corresponding branch cut is located
in the upper half plane.  We depict some of the Riemann sheets in
Fig~\ref{fig:EPs_in_even_sheets}.

\newcommand{\includegraphicsK}[1]{\includegraphics[width=0.28\textwidth]{#1}}

We examine the quasi-momentum in the vicinity of $g=g^{(n)}$ ($n>1$).
A condition of the branch point $\partial_k J(g^{(n)}, k^{(n)})=0$
implies $\partial_g J(g^{(n)}, k^{(n)}) = k^{(n)}/g^{(n)}$ and
$\partial_k^2 J(g^{(n)}, k^{(n)}) = - \pi k^{(n)}/g^{(n)}$. Hence we
have
\begin{equation}
  G^{(2)}(g^{(n)},k^{(n)}) = \frac{2}{\pi}
  .
\end{equation}
From 
\eqref{eq:kIncrement2}, we conclude 
\begin{equation}
  \label{eq:kAroundEP}
\begin{split}
  k_0(g) 
&
  \simeq
  k^{(n)} - \sqrt{\frac{2}{\pi}(g - g^{(n)})}
  ,
\\
  k_n(g) 
&
  \simeq
  k^{(n)} + \sqrt{\frac{2}{\pi}(g - g^{(n)})}
  ,
\end{split}
\end{equation}
where the signs are determined from the numerical results, which are
consistent with the fact $\Im k_0(g) < 0$ and $\Re k_n(g) > 0$ for 
$g < 0$.

For the Riemann surface that consists of $k_n(g)$'s with odd $n$,
the
situation is quite similar to the even $n$ case.  There is a real
branch point, where $k_1(g)$ and $-k_1(g)$, which correspond to an
equivalent eigenstate, degenerate at $(g,k)=(-2/\pi,0)$. A complex
branch point, which we denote $g^{(n)}$ ($n>1$), involves $k_1(g)$ and
$k_n(g)$.  There is no branch point that involves two excited states
at a time.  In the vicinity of $g^{(n)}$, we have
\begin{equation}
  \label{eq:kAroundEPe}
\begin{split}
  k_1(g) 
&
  \simeq
  k^{(n)} - \sqrt{\frac{2}{\pi}(g - g^{(n)})}
  ,
\\
  k_n(g) 
&
  \simeq
  k^{(n)} + \sqrt{\frac{2}{\pi}(g - g^{(n)})}
  ,
\end{split}
\end{equation}
where the signs are determined from the numerical results, and are
consistent with the fact $\Im k_1(g) < 0$ and $\Re k_n(g) > 0$ for 
$g < 0$.

\section{Emulating the eigenenergy anholonomy 
  with complex contour}
\label{sec:energy}
It is straightforward to obtain the Riemann surfaces of eigenenergies
from the analysis of $k_n(g)$ above.  For a given $\bar{k}$, the
quantum number of the center of mass, we have a Riemann surface that
consists of $E_{\bar{k},n}(g)$ for all possible $n$'s.  Only even
(odd) $n$ is possible for even (odd) $\bar{k}$
\cite{Yonezawa-up-20130}.

Let us take an example of the case of $\bar{k} = 0$. The eigenenergies
$E_{0,n}(g)$ with even $n$ form the corresponding Riemann surface.
The branch point $g^{(n)}$, which involves $k_0(g)$ and $k_n(g)$
($n>1$), is introduced as the degeneracy point of $k_{0}(g)$ and
$k_{n}(g)$, as explained above. Also, $g^{(n)}$ is a branch point or
exceptional point for the pair of eigenenergies of $E_{0,0}(g)$ and
$E_{0,n}(g)$. From \eqref{eq:kAroundEP}, we obtain a $\sqrt{}$-type
behavior in the vicinity of $g^{(n)}$:
\begin{equation}
  E_{0,n}(g)-E_{0,0}(g)
  = k^{(n)} \sqrt{\frac{2}{\pi}(g - g^{(n)})}
  +\mathcal{O}(g - g^{(n)})
  .
\end{equation}

\begin{figure}[h]
  \centering
  \includegraphics[width=0.4\textwidth]{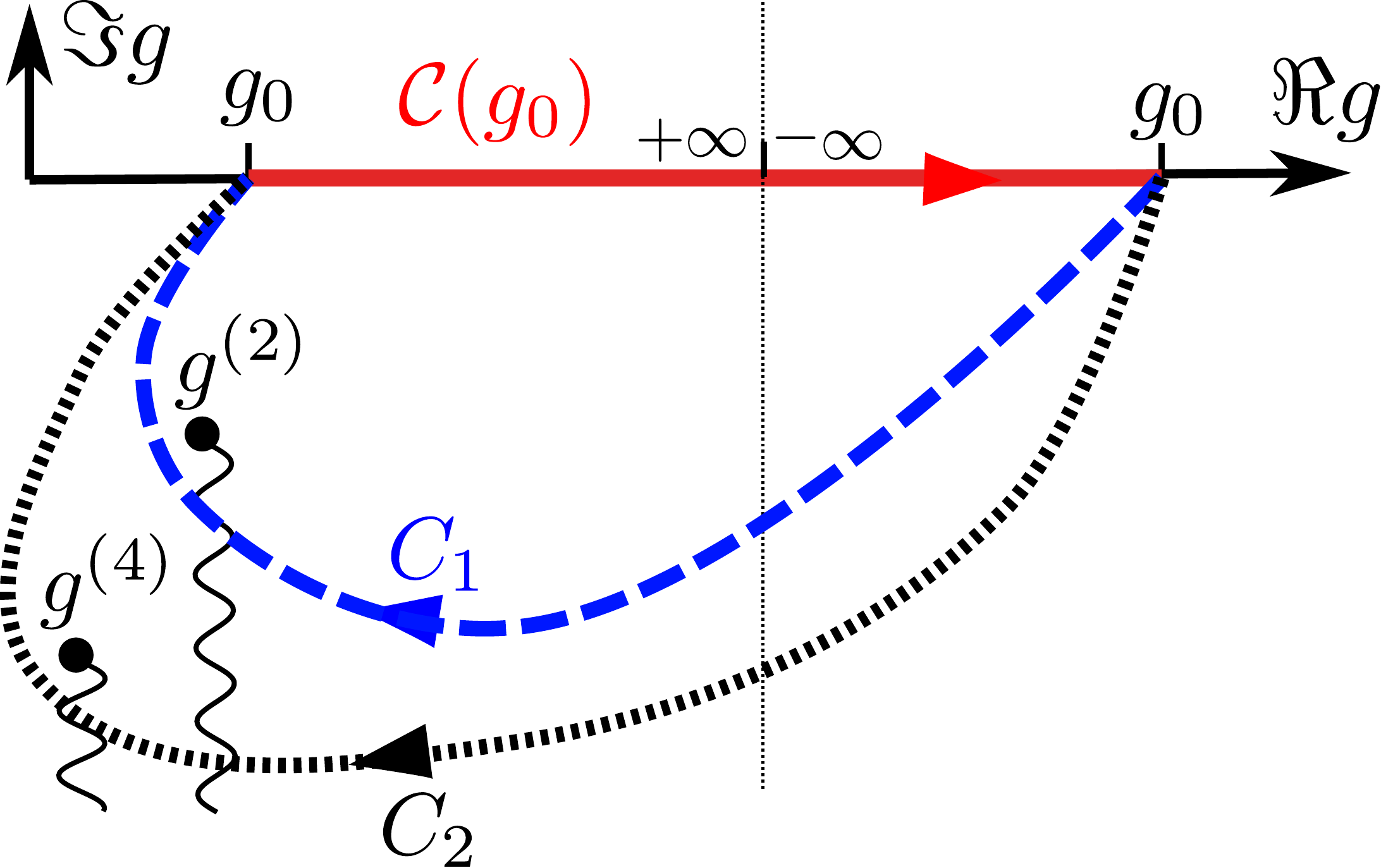}
  \caption{%
    (Color online)
    Schematic picture of contours that
    enclose exceptional points. 
    We depict the real cycle $\CReal(g_0)$ by a thick line.
    Note that $\CReal(g_0)$ traverses the line $g=\pm \infty$.
    The concatenation of $\CReal(g_0)$ and $C_1$ (dashed curve)
    encloses a single exceptional point $\gEP{2}$, and offers 
    the $1$-EP approximation. 
    On the other hand, 
    the concatenation of $\CReal(g_0)$ and $C_2$ (dotted curve)
    encloses two exceptional points $\gEP{2}$ and $\gEP{4}$, 
    and offers the $2$-EP approximation. 
  }
  \label{fig:C1C2}
\end{figure}

Let us explain how we emulate the eigenenergy anholonomy by a closed
contour in the complexified parameter space. More precisely, we
compare the permutations induced by $\CReal(g_0)$ and closed cycles in
the complex $g$-plane (see, figure~\ref{fig:C1C2}).  The real cycle
$\CReal(g_0)$ starts from $g=g_0$ and arrives at a point that is
equivalent to $g_0$, as explained in Sec.~\ref{sec:Ndim}.  Note that
$g$ passes $\pm \infty$ during $\CReal(g_0)$.  On the other hand, the
initial and final points of the closed cycles examined here are $g_0$.
We show that this requires to include all relevant contribution from
the exceptional points (EPs).  We start from ``$N$-EP approximation''
for integer $N$.

Firstly, we examine the contour that encloses only a single
exceptional point $g^{(2)}$ (``$1$-EP approximation''). We explain the
associated parametric evolution 
of
each eigenenergy along the contour.
As for $E_{0,0}(g)$ and $E_{0,2}(g)$, they are exchanged each other
after the completion of the closed cycle. On the other hand, the
closed cycle does not change other eigenenergies.  This is because
$g^{(2)}$ is the branch point 
that involves only
$E_{0,0}(g)$ and
$E_{0,2}(g)$. Hence the resultant permutation is cyclic:
\begin{equation}
  \left(
    E_{0,0}(g_0), E_{0,2}(g_0)
  \right)
  ,
\end{equation}
which mimics the result of $\CReal(g_0)$ only for $E_{0,0}(g_0)$.  We
note that such an approximation breaks down 
when
we repeat the complex
cycle. A similar permutation between $E_{0,0}(g)$ and
$E_{0,n}(g)$ occurs along a closed contour that enclose only $g^{(n)}$
($n>1$). We note that all $1$-EP case involves the ground energy
$E_{0,0}(g)$.

Secondly, we examine a closed contour that encloses only two different
exceptional points, say, $g^{(2)}$ and $g^{(4)}$ (``$2$-EP
approximation'').  The result of the parametric evolution of each
eigenenergy along the closed contour is the following cyclic
permutation:
\begin{equation}
  \left(
    E_{0,0}(g_0), E_{0,2}(g_0), E_{0,4}(g_0)
  \right)
  .
\end{equation}
Hence the result of the single cycle, as for $E_{0,0}(g_0)$ and
$E_{0,2}(g_0)$, mimics the result of $\CReal(g_0)$.  Other cycles that
involve two exceptional points induce a similar permutation.  We note
again that all $2$-EP case involves the ground energy
$E_{0,0}(g)$. This is because all ``elementary'' branch points involve
$E_{0,0}(g)$.

Now it is straightforward to extend our analysis to $N$-EP cases
(``$N$-EP approximation'').  A closed contour that encloses $N$ branch
points $\set{g^{(2m)}}_{m=1}^N$ induces the following cyclic
permutation of $N+1$ eigenenergies:
\begin{equation}
  \left(
    E_{0,0}(g_0), E_{0,2}(g_0), \dots, E_{0,2N}(g_0)
  \right)
  ,
\end{equation}
which approximates the permutation induced by $\CReal(g_0)$, as for
lower-lying eigenenergies.  In this sense, the limit $N\to\infty$
provides us a closed cycle that emulates the eigenenergy anholonomy
induced by $\CReal(g_0)$.

Our analysis suggests that the exceptional points offer ``elements''
of the eigenenergy anholonomy of the two-body Lieb-Liniger model.
This view is an extension of the previous
result~\cite{Kim-PLA-374-1958}.

\section{Eigenspace anholonomy in terms of 
  exceptional points}
\label{sec:eigenspace}

In this section, we explain the role of exceptional points in the
gauge 
theory
that provides a unified formulation of
the phase holonomy and the eigenspace
anholonomy~\cite{Cheon-EPL-85-20001}. In particular, we show an
evaluation of the holonomy matrix $M(C)$ (See, \eqref{eq:Mcovariant}
below), which quantifies the eigenspace anholonomy, using the
exceptional points.  This confirms our view that the exceptional
points constitute a skeleton of the eigenspace anholonomy.  To prepare
this, we make a brief review of the gauge theory of the eigenspace
anholonomy in \S~\ref{sec:gauge_theory}.  We show that each
exceptional point provides a ``local'' contribution to $M(C)$ in
\S~\ref{sec:EP_local}.  By collecting these contributions, we conclude
this section (\S~\ref{sec:EP_global}).

\subsection{Gauge theory of eigenspace anholonomy}
\label{sec:gauge_theory}

We outline the gauge theory of the eigenspace and phase
anholonomies~\cite{Cheon-EPL-85-20001}.  Suppose that the system is
initially in an eigenstate $\ket{\psi_{\bar{k},n}(g)}$ and the
parameter is adiabatically deformed along a cycle $C$.  Let
$\ket{\psi_{\bar{k},n}(g; C)}$ 
denote the final state
induced by the adiabatic time evolution along $C$.  We assume that the
dynamical phase~\cite{Berry-PRSLA-392-45} is removed from
$\ket{\psi_{\bar{k},n}(g; C)}$.  
A simple way to quantify
the eigenspace anholonomy, which concerns about the discrepancy
between $\ket{\psi_{\bar{k},n}(g)}$ and
$\ket{\psi_{\bar{k},n}(g; C)}$, 
is to examine the
overlapping integral or the holonomy matrix 
$\bracket{\psi^{\mathrm{L}}_{\bar{k}'n'}(g)}{\psi_{\bar{k}n}(g;C)}.$
Note that the adiabatic variation of the interaction strength $g$ does
not vary $\bar{k}$, which is the quantum number of the center of mass.
Hence it suffices to focus on the case $\bar{k}'=\bar{k}$:
\begin{equation}
  M_{n',n}(C) 
  \equiv \bracket{\psi^{\mathrm{L}}_{\bar{k} n'}(g)}{\psi_{\bar{k}n}(g;C)}
  ,
\end{equation}
which is independent of $\bar{k}$, as is seen below. Also,
$M_{n',n}(C)$ is non-zero only when the oddness (or evenness) of
$\bar{k}$, $n'$ and $n$ is the same.  Hence $M(C)$ is consist of the
even and odd blocks.

A gauge covariant expression of $M(C)$ is 
\begin{equation}
  \label{eq:Mcovariant}
  M(C)
  =
  \AntiTexp\left(-i\int_C A(g)dg\right)
  \exp\left(i\int_C \AD{}(g)dg\right)
  ,
\end{equation}
where $\AntiTexp$ indicates the anti-path-ordered exponential, and
$A(g)$ and $\AD{}(g)$ are gauge 
connections~\cite{Cheon-EPL-85-20001,Kim-PLA-374-1958}
\begin{equation}
  \begin{split}
  \label{eq:defA}
  A_{n',n}(g)
  &
  \equiv i\bra{\psi^{\mathrm{L}}_{\bar{k} n'}(g)}%
  \left[\partial_g\ket{\psi_{\bar{k},n}(g)}\right]
  \\
  \AD{}_{n',n}(g)
  &
  \equiv \delta_{n',n}A_{\bar{k} n',\bar{k}n}(g)
  ,
  \end{split}
\end{equation}
which are independent of the total momentum $\bar{k}$, too.

Two kinds of gauge invariants are involved in
$M(C)$~\cite{Tanaka-JPA-45-335305}. One is a permutation matrix and
the other is the off-diagonal geometric
phases~\cite{Manini-PRL-85-3067}.

In the following, we impose the parallel transport
condition~\cite{Stone-PRSLA-351-141} for each eigenspace, i.e.,
\begin{equation}
  \label{eq:PTcondition}
  \AD{}_{n',n}(g) = 0. 
\end{equation}
This makes the parametric evolution of the eigenvectors precisely
describe the adiabatic time evolution except the dynamical phase.
Hence it is also suitable to investigate analytic continuation of the
adiabatic parameter for eigenfunction.  Regardless of $C$ being closed
or open, the parametric evolution of eigenvectors is described by the
gauge connection $\AF{}(g)$ as
\begin{equation}
  \label{eq:parametricEvolPsi}
  \ket{\psi_{\bar{k}, n}(g; C)}
  = \sum_{n'} \ket{\psi_{\bar{k}, n'}(g)} 
  \left[\AntiTexp\left(-i\int_C A(g)dg\right)\right]_{n'n}
  .
\end{equation}
In particular, the second factor in \eqref{eq:Mcovariant} vanishes
\begin{equation}
  \label{eq:MPT}
  M(C)
  =
  \AntiTexp\left(-i\int_C A(g)dg\right)
  .
\end{equation}

We obtain the gauge connections from
\eqref{eq:defA},~\eqref{eq:defPsi} and~\eqref{eq:defPsiL}. The
diagonal elements are
\begin{eqnarray}
  &&
  A_{n, n}(g) 
  \nonumber\\ &
  =
  &
    \begin{cases}
    \partial_g
    \left[i\ln\left\{
        \ampg(k)\left(1+\frac{\sin(\pi k)}{\pi k}\right)^{1/2}\right\}\right]
    ,
    &
      \text{%
      for even $n$,
    } 
    \\
    \partial_g
    \left[i\ln\left\{
        \ampe(k)\left(1-\frac{\sin(\pi k)}{\pi k}\right)^{1/2}\right\}\right]
    ,
    &
      \text{%
      for odd $n$
    }
    .
    \end{cases}
\end{eqnarray}
The parallel transport condition~\eqref{eq:PTcondition} implies
\begin{eqnarray}
  \ampg(k) 
  &
  =
  &
  \ampg(0)\sqrt{2}\left(1+\frac{\sin(\pi k)}{\pi k}\right)^{-1/2}
  ,
  \\
  \ampe(k)
  &
  = 
  &
  \alpha\sqrt{2}\left(1-\frac{\sin(\pi k)}{\pi k}\right)^{-1/2}
  ,
\end{eqnarray}
where $\alpha$ is a constant. The off-diagonal elements of the gauge
connections are
\begin{equation}
  A_{n'n}(g)
  =
  -i\frac{4}{\pi}
  \frac{D_{n'}(g)D_{n}(g)}{k'^2-k^2}
  \left(1 - \delta_{n'n}\right)
  ,
\end{equation}
where $k=k_n(g)$ and $k'=k_{n'}(g)$ are assumed, and
\begin{equation}
  \label{eq:D_def}
  D_n(g)
  \equiv
    \begin{cases}
    \left(1 + \frac{\sin(\pi k)}{\pi k}\right)^{-\frac{1}{2}}
    \cos\frac{\pi k}{2}
    ,&
      \text{%
      for even $n$,
    } 
    \\    
    \left(1 - \frac{\sin(\pi k)}{\pi k}\right)^{-\frac{1}{2}}
    {\sin\frac{\pi k}{2}}
    ,&
      \text{%
      for odd $n$.
    }
    \end{cases}
\end{equation}
We will obtain another expression of $D_n(g)$ in 
Appendix~%
\ref{sec:Dfunction}:
\begin{equation}
  \label{eq:Dn_unified}
  D_n(g) 
  = d_n \frac{k_n(g)}{\sqrt{\left\{k_n(g)\right\}^2+g^2 + 2 g/\pi}}
  ,
\end{equation}
where $d_n$ is defined as
\begin{equation}
  \label{eq:dn_def}
  d_n = 
  (-1)^{[n/2]}
  ,
\end{equation}
and
$[x]$ is the maximum integer less than $x$.

\subsection{Contribution from an exceptional point to 
  $M(C)$}
\label{sec:EP_local}

\begin{figure}[h]
  \centering
  \includegraphics[width=0.25\textwidth]{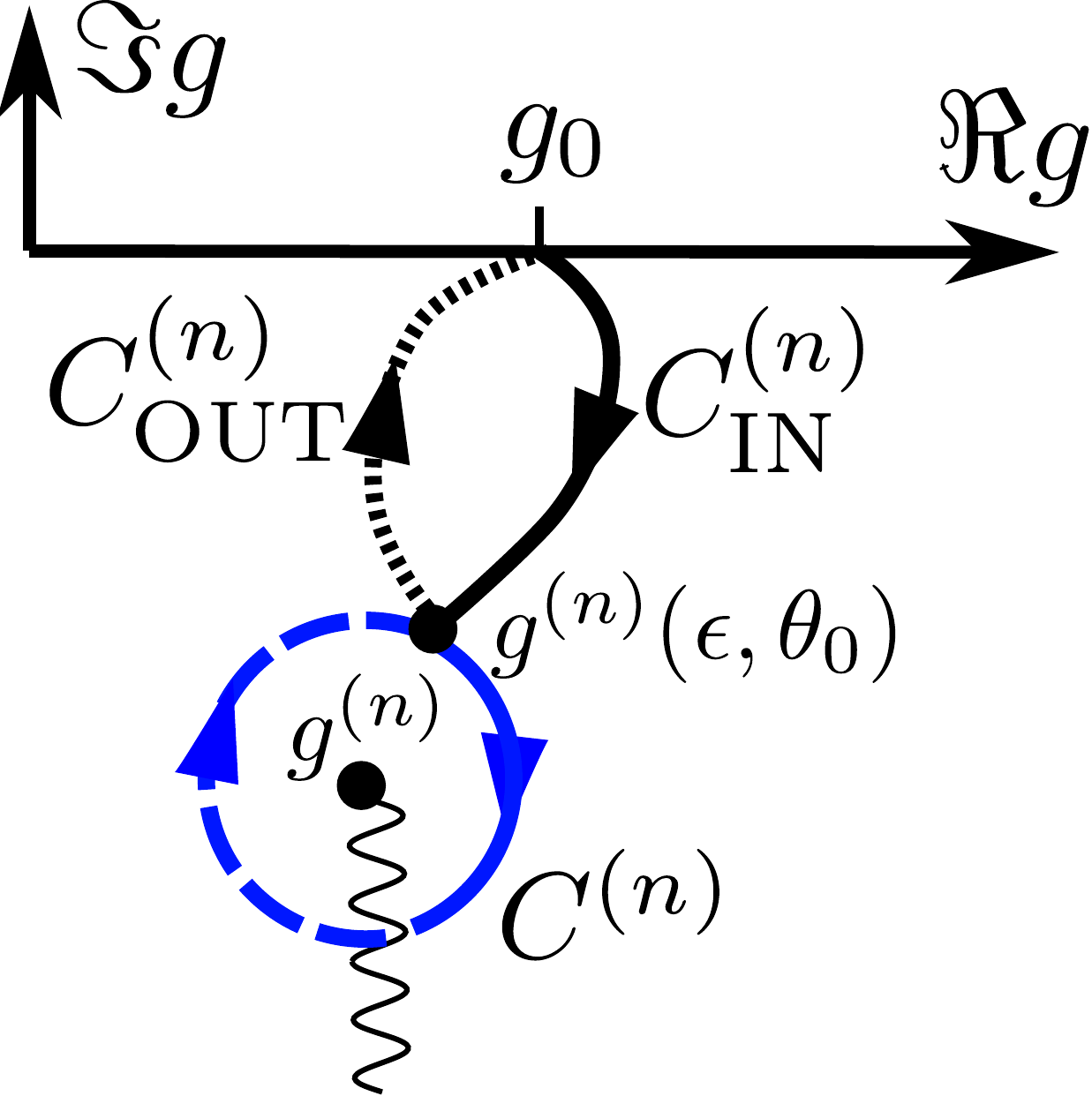}
  \caption{%
    (Color online) Schematic picture of contours for the evaluation of
    \eqref{eq:defMn} and \eqref{eq:EvolveCn}.  The circle $C^{(n)}$
    encircles 
    an
    exceptional point $g^{(n)}$ with a radius $\epsilon$
    (see the main text). The initial point of the cycle $C^{(n)}$ is
    denoted by 
    $g^{(n)}(\epsilon,\theta_0)=g^{(n)}+\epsilon e^{i\theta_0}$.  
    Note that $C^{(n)}$ intersects a branch cut (wavy
    line) that emanates from $g^{(n)}$.  The bold and dotted lines
    that connect $g_0$ and $g^{(n)}(\epsilon,\theta_0)$ are
    $C^{(n)}_{\rm IN}(g_0)$ and $C^{(n)}_{\rm OUT}(g_0)$,
    respectively.  }
  \label{fig:EPn}
\end{figure}

We evaluate $M(C)$ \eqref{eq:MPT}, deforming the integration contour
$C$. The anti-path-ordered exponential 
in $M(C)$ is decomposed into
the contributions
from the exceptional points. Here we focus on the contribution from
the single exceptional point $g^{(n)}$ ($n>1$).  Namely, we will
evaluate the anti-path-ordered exponential of the gauge connection
along the contour 
$\CEP{n} = \set{\gEP{n} + \epsilon e^{i\theta}|\;
  \theta_0 \le \theta \le \theta_0+ 2\pi}$, 
which encircles the
exceptional point $\gEP{n}$ along the clockwise direction with the
radius $\epsilon$ 
(see figure~\ref{fig:EPn}).
In particular, we focus on the limit $\epsilon\downarrow0$, i.e.,
\begin{equation}
  \label{eq:defMn}
  M^{(n)}
  \equiv
  \lim_{\epsilon\downarrow 0}
  \AntiTexp\left(-i\int_{\CEP{n}} \AF{}(g) dg\right)
  ,
\end{equation}
in the following.%
\footnote{%
  The encirclement around an exceptional point has been examined to study 
  the associated phase holonomy~\cite{Arnold-SelectaMathemeticaNewSeries-1-1,Mailybaev-PRA-72-014104,Dietz-PRL-106-150403}.}

First, we show that the gauge connection $\AF{}(g)$ is singular at the
exceptional point.  Assume that $\epsilon\equiv g - g^{(n)}$ is small.
As explained in Sec.~\ref{sec:wavenumber}, the quasi-momenta
$k_{\nb}(g)$ and $k_n(g)$ are degenerate at $\epsilon=0$, i.e.,
$g=g^{(n)}$, where $\nb=0$ for even $n$, and $\nb=1$ for odd $n$.  In
Appendix~%
\ref{sec:DDaroundEP}, we show that
\begin{equation}
  \label{eq:DD_expand}
  D_n(g)D_{\nb}(g) 
  = d_n \frac{i \pi^{1/2}k^{(n)}}{2^{3/2}\epsilon^{1/2}}
  \left[1+\mathcal{O}(\epsilon^{1/2})\right]
  .
\end{equation}
On the other hand, 
\eqref{eq:kAroundEP} 
implies 
\begin{equation}
  \left(\left[k_n(g)\right]^2 - \left[k_{\nb}(g)\right]^2 \right)^{-1}
  = 
  \frac{\pi^{1/2}}{2^{5/2} k^{(n)} \epsilon^{1/2}}
  \left[1+\mathcal{O}(\epsilon^{1/2})\right]
  .
\end{equation}
Combining these factors, each of which diverges
$\Order(\epsilon^{-1/2})$ as $\epsilon\to0$, we find
\begin{equation}
  A_{n,\nb}(g)
  = d_n \frac{1}{4\epsilon}  \left[1+\mathcal{O}(\epsilon^{1/2})\right]
  .
\end{equation}
Accordingly, the gauge connection within the subspace spaned by
$\nb$-th and $n$-th eigenstates is
\begin{equation}
  \label{eq:A2}
  \left[
    \begin{array}{cc} 
      A_{\nb,\nb}(g)& A_{\nb,n}(g)\\
      A_{n,\nb}(g)& A_{n_0,n_0}(g)
    \end{array}
  \right]
  = 
  i\frac{d_n}{\epsilon} R \left[1+\mathcal{O}(\epsilon^{1/2})\right]
  ,
\end{equation}
where
\begin{equation}
  \label{eq:defR}
  R
  \equiv
  -\frac{1}{4}
  \left[
    \begin{array}{cc} 
    0& -i\\
    i& 0
    \end{array}
  \right]
  .
\end{equation}
The leading term is proportional to $\epsilon^{-1}$ and single-valued
around $\epsilon=0$.  The next leading term exhibits weaker divergence
$\Order(\epsilon^{-1/2})$ and multiple-valuedness. Other matrix
elements of \eqref{eq:defA} exhibit, at most, the weaker divergence
$\Order(\epsilon^{-1/2})$.

We examine $M^{(2)}$ by expanding the anti-path-ordered exponential in
\eqref{eq:defMn}:
\begin{eqnarray}
  M^{(2)}
  &
  = 
  &
  \lim_{\epsilon\downarrow 0}
  \Pantiorder
  \left[
    1 - i\int_{\CEP{2}}A(g) dg
  \right.\nonumber \\ &{}& \qquad\qquad \left.
    - \frac{1}{2}\int_{\CEP{2}}A(g_1) dg_1\int_{\CEP{2}}A(g_2) dg_2
  \right.\nonumber \\ &{}& \qquad\qquad \left.
    +\dots\right]
  ,
\end{eqnarray}
where $\Pantiorder$ indicates the anti-path ordering product of
matrices.  From the argument above, the dominant part of the gauge
connection has a block-diagonal structure
\begin{equation}
  \label{eq:AF2}
  \AF{}(g^{(2)}+\epsilon)
  = 
  - \frac{i}{{\epsilon}}
  \left[
    \begin{array}{c|c}
      R& 0\\ \hline
      0& 0\\
    \end{array}
  \right]
  + \mathcal{O}(\epsilon^{-\frac{1}{2}})
  ,
\end{equation}
where the matrix representation involves only the subspace that
consists of $\ket{\psi_{0, n} (g)}$ with even $n$($\ge 0$).  Since the
circumference of $\CEP{2}$ is $2\pi\epsilon$, only the leading term in
\eqref{eq:AF2} contributes to $M^{(2)}$:
\begin{equation}
  M^{(2)}
  =
  \left[
    \begin{array}{c|c}
      \exp\left(-i 2\pi R\right)& 0\\ \hline
      0& 1\\
    \end{array}
  \right]
  .
\end{equation}
It is straightforward to see
\begin{equation}
  \exp\left(-i 2\pi R\right)
  =
  \left[
    \begin{array}{cc}
    0& -1\\
    1&  0
    \end{array}
  \right]
\end{equation}
from 
\eqref{eq:defR}. Hence we obtain
\begin{equation}
  \label{eq:M2}
  M^{(2)}
  =
  \left[
    \begin{array}{cc|c}
      0& -1& 0\\
      1&  0& 0\\ \hline
      0&  0& 1
    \end{array}
  \right]
  ,
\end{equation}
which describes the parametric evolution of eigenvectors along
$\CEP{2}$. The bound state $\ket{\psi_{0, 0} (g)}$ evolves into
$\ket{\psi_{0, 2}(g)}$. The partner $\ket{\psi_{0, 2}(g)}$ evolves
into $-\ket{\psi_{0, 0} (g)}$, where an extra phase factor $(-1)$ is
acquired. Other eigenvectors are remain unchanged.

It is straightforward to obtain $M^{(n)}$ for an arbitrary $n$($>1$)
\begin{equation}
  \label{eq:Mn}
  \begin{split}
  M^{(n)}_{n''n'}
  &
  = d_n\left(\delta_{n''\nb}\delta_{n'n}-\delta_{n''n}\delta_{n'\nb}\right)
  \\&\qquad
  + (1-\delta_{n''\nb})(1-\delta_{n''n})\delta_{n''n'}
  .
  \end{split}
\end{equation}
We depict a few of them:
\begin{equation}
  M^{(4)}
  =
  \left[
    \begin{array}{ccc|c}
      0&  0& 1& 0\\
      0&  1& 0& 0\\
      -1& 0& 0& 0\\ \hline
      0&  0& 0& 1
    \end{array}
  \right]
  ,\qquad
  M^{(6)}
  =
  \left[
    \begin{array}{cccc|c}
      0&  0& 0& -1&0\\
      0&  1& 0& 0&0\\
      0&  0& 1& 0&0\\ 
      1&  0& 0& 0&0\\ \hline
      0&  0& 0& 0& 1
    \end{array}
  \right]
  .
\end{equation}

\subsection{Combining multiple-EP contributions}
\label{sec:EP_global}

Let us examine the analytic continuation of eigenvector 
$\ket{\psi_{0, n'}(g_0)}$, where $g_0$ is real and
$n'$ is even.
We extend $\ket{\psi_{0, n'}(g_0)}$
along 
the cycle (see figure~\ref{fig:EPn})
\begin{equation}
 C^{(n)}(g_0)
 = C^{(n)}_{\mathrm{IN}}(g_0)\cat{}C^{(n)}_{\mathrm{EP}}\cat{}
 C^{(n)}_{\mathrm{OUT}}(g_0) 
\end{equation}
($n>1$), where $a\circ b$ indicates the concatenation of 
paths $a$ and $b$.
We note that all eigenvectors remain unchanged
against the parametric evolution along
$C^{(n)}_{\mathrm{IO}}(g_0)\equiv
C^{(n)}_{\mathrm{IN}}(g_0)\cat 
C^{(n)}_{\mathrm{OUT}}(g_0)$, because
$C^{(n)}_{\mathrm{IO}}(g_0)$ encloses no branch point (figure~\ref{fig:EPn}).
Hence,
from 
\eqref{eq:parametricEvolPsi},
$\ket{\psi_{0, n}(g_0; C^{(n)}(g_0))}$ has the following expression
\begin{equation}
  \label{eq:EvolveCn}
  \ket{\psi_{0, n}(g_0; C^{(n)}(g_0))}
  = \sum_{n''} \ket{\psi_{0, n''}(g_0)} M^{(n)}_{n'',n'}
  .
\end{equation}

Next, we consider the effect of two exceptional points $\gEP{2}$ and
$\gEP{4}$ using a contour $C^{(2,4)}\equiv C^{(2)}(g_0)\cat
C^{(4)}(g_0)$ (see, figures~\ref{fig:C1C2} and~\ref{fig:EPn}).  To
carry out this, let us extend the eigenvectors in \eqref{eq:EvolveCn}
with $n=2$ along $C^{(4)}(g_0)$. We find
\begin{equation}
  \label{eq:EvolveC24}
  \ket{\psi_{0, n}(g_0; C^{(2,4)}(g_0))}
  = \sum_{n''} \ket{\psi_{0, n''}(g_0; C^{(4)}(g_0))} M^{(2)}_{n'',n'}
  .
\end{equation}
Hence we obtain, using 
\eqref{eq:EvolveCn} with $n=4$,
\begin{equation}
  \ket{\psi_{0, n}(g_0; C^{(2,4)}(g_0))}
  = \sum_{n''} \ket{\psi_{0, n''}(g_0)} M^{(2,4)}_{n'',n'}
  ,
\end{equation}
where
\begin{equation}
  M^{(2,4)} = M^{(4)}M^{(2)}
  .
\end{equation}

Furthermore, we examine the analytic continuation of
$\ket{\psi_{0, n}(g_0)}$ along
\begin{equation}
 C^{(2,\dots,2m)}(g_0)
 \equiv
 C^{(2,\dots,2m-2)}(g_0)\cat C^{(2m)}(g_0) 
\end{equation}
($m>1$).
In a similar way above, we find
\begin{equation}
  \ket{\psi_{0, n}(g_0; C^{(2,\dots,2m)}(g_0))}
  = \sum_{n''} \ket{\psi_{0, n''}(g_0)} M^{(2,\dots,2m)}_{n'',n'}
  ,
\end{equation}
where
\begin{equation}
  \label{eq:Mrecursion}
  M^{(2,\dots,2m)} = M^{(2m)}M^{(2,\dots,2m-2)}
  .
\end{equation}
As shown in 
Appendix~%
\ref{sec:M2dots2m}, we obtain
\begin{eqnarray}
  \label{eq:M2dots2m}
  \lefteqn{
    M^{(2,\dots,2m)}_{n'',n'}
  }\nonumber\\ 
  &
  =
  &
  \sum_{m'=0}^{m-1}\delta_{n'',2(m'+1)}\delta_{n',2m'}
  + (-1)^m \delta_{n'',0}\delta_{n',2m}
  \nonumber\\&{}&\qquad
  + \prod_{m'=0}^{m}\left(1-\delta_{n'',2m'}\right)\delta_{n'',n'}
  ,
\end{eqnarray}
where the first and second term describe the shift of eigenstates
$(0,\dots,2m-2)\mapsto(2,\dots,2m)$, and the shift of eigenstate
$2m\mapsto 0$ with a phase factor $(-1)^m$.  The other eigenstates
remain unchanged by the closed contour $C^{(2,\dots,2m)}(g_0)$.  Hence
this contour accurately describes the shift of eigenstates up to
$(2m-2)$-th excited 
states.
In this sense, $C^{(2,\dots,2m)}(g_0)$
emulates $C$ in the limit $m\to\infty$ as for the exotic quantum
holonomy induced by $\CReal(g_0)$ along the real axis of $g$, i.e.,
\begin{equation}
  M^{(2,\dots,2m)}\to M\left(\CReal(g_0)\right)
\end{equation}
as $m\to\infty$.

\section{Discussion}
\label{sec:discussion}
First, we compare the present result with
Ref.~\cite{Kim-PLA-374-1958}, where the correspondence between the
exotic quantum holonomy and exceptional points is examined in families
of quantum kicked spin-$\frac{1}{2}$. First of all, because a kicked
spin is a periodically driven system, the exotic quantum holonomy of
the eigenvalues and the eigenvectors of Floquet operator, which is the
time evolution operator during the period of a driving force, is
investigated. Hence the physical context of the exotic quantum
holonomy is slightly different from the one in autonomous systems. On
the other hand, these two models have the same relationship between
the quantum holonomy and the exceptional points, as a whole.  For
example, the multiple-valuedness of eigenvalues and eigenvectors is
governed by the exceptional points. The non-Abelian gauge connection
has a $\epsilon^{-1}$-divergence around 
an
exceptional point (see
\eqref{eq:A2}), where $\epsilon$ is the distance from the exceptional
point in the parameter space.  This divergence comprises the
permutation of eigenvectors against a tiny loop around the exceptional
point (see \eqref{eq:M2}).  However, we find a subtle difference on
the analyticity of the gauge connection. As for the kicked spin, the
gauge connection is single-valued in the parameter space.  We may say
that the gauge connection has a degree-$1$ pole at an exceptional
point of the kicked spin. On the other hand, as for the gauge
connection of two-body Lied-Liniger model, an exceptional point is not
only a divergent point, but also a branch point. However the
multiple-valuedness appears only in the higher-order correction terms
about $\epsilon$ (see \eqref{eq:A2}).

Second, it is certain that we should see whether the present
observations apply to Lieb-Liniger model with an arbitrary number of
particles, as we focus on the two-body case.  We may expect that a
similar scenario on the interplay of the exotic quantum holonomy and
the exceptional points can be applicable.  For example, according to
the strong coupling expansion explained in
Section~\ref{sec:wavenumber}, the degree of exceptional points is $2$
and each exceptional point connects the ground state and an excited
state, regardless of the number of
particles~\cite{Ushveridze-JPA-21-955}.  On the other hand, however,
there remain subtle points. For example, as for two-body case, the
repetitions of the adiabatic cycle $\CReal(g_0)$ and its inverse
connect all eigenstates once we specify the total momentum. We call
the collection of such eigenstates a family~\cite{Yonezawa-up-20130}.
From the present analysis, a family corresponds to a Riemann surface
of eigenenergy. The analytic continuation of the interaction strength
$g$ can connect any pair of eigenstates in a family.  However, as
discovered in Ref.~\cite{Yonezawa-up-20130}, there is an infinite
number of families in three-body Lieb-Liniger model.  For now, whether
or not a family corresponds to a Riemann surface of eigenenergy is
unknown, because several families might be connected in a region far
from the real axis of a Riemann surface.  In other words, the question
is open as to whether there is any exceptional point that is
``inaccessible'' by the real cycles.  Suppose that there is no such
inaccessible exceptional point. This implies one-to-one correspondence
between a family and a Riemann surface.  Although this might suggest
that the exceptional point picture obtained for the two body case is
applicable to an arbitrary number of particles, another question is
raised.  There is only a single Riemann surface for a given total
momentum when the number of particles is two.  The number of Riemann
surfaces, however, is infinite for the number of particle is three.
We do not know how such a proliferation of Riemann surfaces against
the increment of the number of particles is possible.

\section{Summary}
\label{sec:summary}
We have shown the direct link between the exotic quantum holonomy in
eigenenergies and eigenspaces, and the exceptional points, which are
degeneracy points in the complexified parameter space in two-body
Lieb-Liniger model.  With the help of Bethe ansatz, we examine the
Riemann surface of quasi-momentum. All exceptional points in the lower
half plane participate the eigenenergy anholonomy. Also the
non-Abelian gauge connection introduced for the eigenspace anholonomy
exhibits divergent behavior around the exceptional point as well as
tiny multiple-valuedness correction. The exceptional points offer
building blocks of the eigenspace anholonomy.  It remains to be seen
how the current 
result
is to be extended to systems with an arbitrary number
of particles.

\section*{Acknowledgments}
AT wishes to thank Satoshi Ohya for discussion.  This work has
been partially supported by the Grant-in-Aid for Scientific Research
of MEXT, Japan (Grant numbers 22540396 and 21540402).

\appendix
\section{$D$-function}
\label{sec:Dfunction}

We show how we obtain \eqref{eq:Dn_unified}.  In the following, we
assume that $n$ is even.  It is straightforward to obtain a similar
argument for odd $n$.  Using the fact that $k=k_n(g)$ satisfies
\eqref{eq:Bethe21g}, we find that the numerator and the denominator of
\eqref{eq:D_def} satisfy
\begin{equation}
  \cos\frac{z\pi}{2}
  = \pm\sqrt{\frac{k^2}{k^2+g^2}}
  ,
\end{equation}
and
\begin{equation}
  \sqrt{1 + \frac{\sin(\pi k)}{\pi k}}
  = \sqrt{\frac{\pi(k^2+g^2)+2g}{\pi(k^2+g^2)}}
  ,
\end{equation}
respectively.
Hence we obtain
\begin{equation}
  D_n(g) 
  = \pm \frac{k_n(g)}{\sqrt{\{k_n(g)\}^2+g^2 + 2 g/\pi}}
  .
\end{equation}
There remains the ambiguity of sign. It is chosen so as to be
consistent with the behavior of \eqref{eq:D_def} in the real axes:
\begin{equation}
  D_n(g) 
  = (-1)^{n/2} \frac{k_n(g)}{\sqrt{\{k_n(g)\}^2+g^2 + 2 g/\pi}}
  .
\end{equation}
Hence we obtain \eqref{eq:Dn_unified} in the main text.

\section{A derivation of 
  \protect\eqref{eq:DD_expand}}
\label{sec:DDaroundEP}
We examine the singular behavior of the gauge connection
\eqref{eq:defA} around the exceptional point $g^{(n)}$ ($n>1$). We
assume that $\epsilon\equiv g - g^{(n)}$ is small.  Two quasi-momenta
$k_n(g)$ and $k_{\nb}(g)$ degenerates at $(\epsilon,k) = (0,
k^{(n)})$, where $\nb = 0$ for even $n$, and $\nb = 1$ for odd $n$.
We summarize \eqref{eq:kAroundEP} and~\eqref{eq:kAroundEPe}
\begin{equation}
  \begin{split}
    k_n(g)
    &
    =
    k^{(n)}+\left(\frac{2}{\pi}\epsilon\right)^{1/2}+\mathcal{O}(\epsilon)
    .\\
    k_{\nb}(g)
    &
    =
    k^{(n)}
    -\left(\frac{2}{\pi}\epsilon\right)^{1/2}+\mathcal{O}(\epsilon)
  .
  \end{split}
\end{equation}
We choose that the branch cut emanating from $g^{(n)}$ is parallel to
the imaginary axis, and is confined within the lower half plane.
Namely, we suppose that
\begin{equation}
  -\frac{\pi}{2} < \Arg\epsilon \le \frac{3\pi}{2}
\end{equation}
in the Riemann sheet where we are working.

We will examine
\begin{equation}
  r_n(g)
  \equiv
  \left\{k_n(g)\right\}^2+g^2+2g/\pi
\end{equation}
in order to evaluate $D_n(g)= d_n k_n(g)/\sqrt{r_n(g)}$ around $g^{(n)}$.
We start from the real axis, where $\Arg r_n(g)=0$ holds.
Hence we expect
\begin{equation}
  \label{eq:Arg_r_n_at_the_real_axis}
  -\pi < \Arg r_n(g)\le \pi
\end{equation}
holds around the region between the real axis and $g^{(n)}$, as $\Arg
r_n(g)$ has no singular point there.  We examine $\Arg\; r_{\nb}(g)$
in a similar way.  As for real $g$, we have $r_{\nb}(g) >0$ for $g> -
(2/\pi)\nb$ and $r_{\nb}(g) <0$ for $g<- (2/\pi)\nb$.  Note that
$\sqrt{r_{\nb}(g)}$ has a branch point at $g=0$, and the corresponding
branch cut locates at the imaginary axis within the upper half
plane. So we choose $\Arg r_{\nb}(g) = 0$ for $g>- (2/\pi)\nb$ and
$\Arg r_{\nb}(g) = -\pi$ for $g<- (2/\pi)\nb$.  Hence we expect that
\begin{equation}
  \label{eq:Arg_r_0_at_the_real_axis}
  -\frac{3\pi}{2} <  \Arg r_{\nb}(g) \le \frac{\pi}{2}
\end{equation}
is valid in the region between the real axis and $g^{(n)}$.

We move to the vicinity of the exceptional point $g^{(n)}$, where we
obtain
\begin{eqnarray}
    r_n(g)
    &
    = 
    &
    \frac{2^{3/2}}{\pi^{1/2}} k^{(n)} \epsilon^{1/2}
    \left[1+\mathcal{O}(\epsilon^{1/2})\right]
    ,\\
    r_{\nb}(g)
    &
    =
    &
    -\frac{2^{3/2}}{\pi^{1/2}} k^{(n)} \epsilon^{1/2}
    \left[1+\mathcal{O}(\epsilon^{1/2})\right]
    .
\end{eqnarray}
Note that $\Arg\; r_n(g)$ and $\Arg\; r_{\nb}(g)$ are singular at
$g^{(n)}$, i.e., $\epsilon=0$.  From $g^{(n)}$, the real axis is
located in the direction $\Arg\epsilon = \pi/2$, where
\eqref{eq:Arg_r_n_at_the_real_axis} and
~\eqref{eq:Arg_r_0_at_the_real_axis} are expected to be valid.  We
choose $\Arg k^{(n)}$ so as to satisfy
\begin{equation}
  \label{eq:Arg_kn_condition}
  -\pi < \frac{\pi}{4}+\Arg k^{(n)} \le \pi
  ,
\end{equation}
in order to be consistent with \eqref{eq:Arg_r_n_at_the_real_axis}.
On the other hand, there remains ambiguity of $\Arg r_{\nb}(g) =
\Arg(-1) + \frac{\pi}{4} + \Arg k^{(n)}$.  We resolve this using
\eqref{eq:Arg_kn_condition} and
\eqref{eq:Arg_r_0_at_the_real_axis}. We conclude
\begin{equation}
  \label{eq:Arg_r_0_at_EP}
  \Arg r_{\nb}(g) = -\pi + \frac{1}{2}\Arg \epsilon + \Arg k^{(n)}
  ,
\end{equation}
which implies 
\eqref{eq:DD_expand} in the main text.

\section{A proof of 
  \protect\eqref{eq:M2dots2m}}
\label{sec:M2dots2m}

We 
prove 
\eqref{eq:M2dots2m} 
by
induction.  Note that, as
for $m=1$, \eqref{eq:M2} implies Eq.~\eqref{eq:M2dots2m}.  Hence it
suffices to prove \eqref{eq:M2dots2m} for $m=m'+1$ using the
assumption that \eqref{eq:M2dots2m} holds for $m=m'$($>1$).  For
simplicity, $m'$ is denoted by $m$.  From the recursion relation
\eqref{eq:Mrecursion}, we have
\begin{equation}
  M^{(2,\dots,2m+2)}_{n'',n'} 
  = \sum_{n}M^{(2m+2)}_{n''n}M^{(2,\dots,2m)}_{nn'}
  .
\end{equation}
We find from 
\eqref{eq:Mn}
\begin{eqnarray}
  M^{(2,\dots,2m+2)}_{n'',n'} 
  &
  =
  &
  (-1)^{m+1}\delta_{n''0}M^{(2,\dots,2m)}_{2(m+1),n'}
  \nonumber\\&{}&\qquad
  + (-1)^{m}\delta_{n'',2(m+1)}M^{(2,\dots,2m)}_{0n'}
  \nonumber\\&{}&\qquad
  + (1-\delta_{n''0})(1-\delta_{n''2(m+1)})
  M^{(2,\dots,2m)}_{n''n'}
  .
  \nonumber\\
\end{eqnarray}
Hence, 
\eqref{eq:M2dots2m} implies
\begin{eqnarray}
  M^{(2,\dots,2m+2)}_{n'',n'}
  &
  =
  &
  \sum_{m'=0}^{m}\delta_{n'',2(m'+1)}\delta_{n',2m'}
  \nonumber\\&{}&\qquad
  + (-1)^{m+1} \delta_{n'',0}\delta_{n',2(m+1)}
  \nonumber\\&{}&\qquad
  + \left[\prod_{m'=0}^{m+1}\left(1-\delta_{n'',2m'}\right)\right]
  \delta_{n'',n'}
  .
  \nonumber\\
\end{eqnarray}
Hence the proof of 
\eqref{eq:M2dots2m} is completed.


\begin{thebibliography}{10}
\expandafter\ifx\csname url\endcsname\relax
  \def\url#1{{\tt #1}}\fi
\expandafter\ifx\csname urlprefix\endcsname\relax\def\urlprefix{URL }\fi
\providecommand{\eprint}[2][]{\url{#2}}

\bibitem{Born-ZP-51-165}
Born M and Fock V 1928 {\it Z. Phys.\/} {\bf 51} 165

\bibitem{Kato-JPSJ-5-435}
Kato T 1950 {\it J. Phys. Soc. Japan\/} {\bf 5} 435

\bibitem{LonguetHiggins-PRSL-344-147}
Longuet-Higgins H~C 1975 {\it Proc. R. Soc. London\/} {\bf A 344} 147

\bibitem{Mead-JCP-70-2284}
Mead C and Truhlar D~G 1979 {\it J. Chem. Phys.\/} {\bf 70} 2284

\bibitem{Berry-PRSLA-392-45}
Berry M~V 1984 {\it Proc. R. Soc. London\/} {\bf A 392} 45

\bibitem{Wilczek-PRL-52-2111}
Wilczek F and Zee A 1984 {\it Phys. Rev. Lett.\/} {\bf 52} 2111

\bibitem{Bohm-GPQS-2003}
Bohm A, Mostafazadeh A, Koizumi H, Niu Q and Zwanziger Z 2003 
{\it The Geometric Phase in Quantum Systems\/} (Berlin: Springer)

\bibitem{Cheon-PLA-248-285}
Cheon T 1998 {\it Phys. Lett. A\/} {\bf 248} 285

\bibitem{Tanaka-PRL-98-160407}
Tanaka A and Miyamoto M 2007 {\it Phys. Rev. Lett.\/} {\bf 98}
  160407

\bibitem{Kim-PLA-374-1958}
Kim S~W, Cheon T and Tanaka A 2010 {\it Phys. Lett. A\/} {\bf 374} 1958

\bibitem{KatoExceptionalPoint}
Kato T 1980 {\it Perturbation Theory for Linear Operators\/} (Berlin:
  Springer-Verlag) chap~II corrected printing of the second ed

\bibitem{Heiss-JMP-32-3003}
Heiss W~D and Steeb W~H 1991 {\it J. Math. Phys.\/} {\bf 32} 3003

\bibitem{Heiss-CzecJP-54-1091}
Heiss W 2004 {\it Czechoslovak Journal of Physics\/} {\bf 54} 1091

\bibitem{phhqp}
See, e.g., Bender C, Fring A, G\"unther U and Jones H 2012
{``Quantum physics with non-Hermitian operator''}, 
{\it J. Phys. A: Math. Theor.} {\bf 45} 440301

\bibitem{biorthogonal}
See, e.g., 
Moiseyev N 2011 {\it Non-Hermitian Quantum Mechanics} 
(New York: Cambridge Univ. Press)

\bibitem{Lieb-PR-130-15}
Lieb E~H and Liniger W 1963 {\it Phys. Rev.\/} {\bf 130} 1605

\bibitem{Ushveridze-JPA-21-955}
Ushveridze A~G 1988 {\it J. Phys. A.\/} {\bf 21} 955

\bibitem{Yonezawa-up-20130}
Yonezawa N, Tanaka A and Cheon T 2013 
{\it Phys. Rev. A} {\bf 87} 062113

\bibitem{Duerr-PRA-79-023614}
D{\"u}rr S, Garc{\'ia}-Ripoll J~J, Syassen N, Bauer D~M, Lettner M, Cirac J~I
  and Rempe G 2009 {\it Phys. Rev. A\/} {\bf 79} 023614

\bibitem{Olshanii-PRL-81-938}
Olshanii M 1998 {\it Phys. Rev. Lett.\/} {\bf 81} 938

\bibitem{Haller-Science-325-1224}
Haller E, Gustavsson M, Mark M~J, Danzl J~G, Hart R, Pupillo G and N{\"a}gerl
  H~C 2009 {\it Science\/} {\bf 325} 1224

\bibitem{Haller-PRL-104-153203}
Haller E, Mark M~J, Hart R, Danzl J~G, Reichs{\"o}llner L, Melezhik V,
  Schmelcher P and N{\"a}gerl H~C 2010 {\it Phys. Rev. Lett.\/} {\bf 104}
  153203

\bibitem{Cheon-EPL-85-20001}
Cheon T and Tanaka A 2009 {\it Europhys. Lett.\/} {\bf 85} 20001

\bibitem{Tanaka-JPA-45-335305}
Tanaka A, Cheon T and Kim S~W 2012 {\it J. Phys. A: Math. Theor.}
{\bf 45} 335305

\bibitem{Manini-PRL-85-3067}
Manini N and Pistolesi F 2000 {\it Phys. Rev. Lett.\/} {\bf 85} 3067

\bibitem{Stone-PRSLA-351-141}
Stone A~J 1976 {\it Proc. R. Soc. London\/} {\bf A 351} 141

\bibitem{Arnold-SelectaMathemeticaNewSeries-1-1}
Arnold  V~I 1991 {\it Selecta Mathematica, New Series} {\bf 1} 1

\bibitem{Mailybaev-PRA-72-014104}
Mailybaev A~A. Kirillov O N, and Seyranian A~P 2005, 
{\it Phys. Rev. A} {\bf 72} 014104

\bibitem{Dietz-PRL-106-150403}
Dietz B, Harney H L, Kirillov O N, Miski-Oglu M, Richter A 
and Sch{\"a}fer F 2011,
{\it Phys. Rev. Lett.} {\bf 106} 150403

\end{thebibliography}

\providecommand{\newblock}{}


\end{document}